\documentclass[a4paper,aps,pre,superscriptaddress,floatfix,nofootinbib,twocolumn,longbibliography]{revtex4-1}
\usepackage{bbm}
\usepackage{bbold}
\usepackage{bbm}
\usepackage[pdftex]{graphicx}
\usepackage{latexsym,amsmath,verbatim,amssymb,txfonts}
\usepackage{lipsum}
\usepackage{color}
\usepackage{rotating}
\usepackage{verbatim}
\usepackage{multirow}
\usepackage[english]{babel}
\usepackage{comment}
\usepackage{bm}
\usepackage{amsmath}
\usepackage{amsfonts}
\usepackage{amssymb}
\usepackage{float}
\usepackage{color}
\usepackage{braket}
\usepackage{blkarray, bigstrut}

\DeclareMathOperator{\lap}{\mathbf{L}}
\DeclareMathOperator{\bnd}{\mathbf{B}}

\newcommand\topstrut[1][1.2ex]{\setlength\bigstrutjot{#1}{\bigstrut[t]}}
\newcommand\botstrut[1][0.9ex]{\setlength\bigstrutjot{#1}{\bigstrut[b]}}

\definecolor{internationalkleinblue}{rgb}{0.0, 0.18, 0.65}
\definecolor{brickred}{rgb}{0.8, 0.25, 0.33}
\definecolor{amber}{rgb}{1.0, 0.75, 0.0}
\definecolor{applegreen}{rgb}{0.55, 0.71, 0.0}
\definecolor{darkviolet}{rgb}{0.58, 0.0, 0.83}

\begin{document}
\title{Diffusion-driven instability of topological signals coupled by the Dirac operator}

\author{Lorenzo Giambagli} 
\email{lorenzo.giambagli@gmail.it}
\affiliation{Department of Physics and Astronomy, University of Florence, INFN \& CSDC, Sesto Fiorentino, Italy}
\affiliation{Department of Mathematics \& naXys, Namur Institute for Complex Systems, University of Namur, Rue Grafé 2, B5000 Namur, Belgium}

\author{Lucille Calmon}
\thanks{These two authors contributed equally}
\affiliation{School of Mathematical Sciences, Queen Mary University of London, London E1 4NS, UK}

\author{Riccardo Muolo}
\thanks{These two authors contributed equally}
\affiliation{Department of Mathematics \& naXys, Namur Institute for Complex Systems, University of Namur, Rue Grafé 2, B5000 Namur, Belgium}
\affiliation{Department of Applied Mathematics, Mathematical Institute Federal University of Rio de Janeiro, Avenida
Athos da Silveira Ramos, 149, Rio de Janeiro 21941-909, Brazil}

\author{Timoteo Carletti}
\affiliation{Department of Mathematics \& naXys, Namur Institute for Complex Systems, University of Namur, Rue Grafé 2, B5000 Namur, Belgium}

\author{Ginestra Bianconi}
\email{ginestra.bianconi@gmail.com}
\affiliation{School of Mathematical Sciences, Queen Mary University of London, London E1 4NS, UK}
\affiliation{The  Alan  Turing  Institute,  96  Euston  Road,  London,  NW1  2DB,  United  Kingdom}

\date{\today}

\begin{abstract}
The study of reaction-diffusion systems on networks is of paramount relevance for the understanding of nonlinear processes in systems where the topology is intrinsically discrete, such as the brain. Until now reaction-diffusion systems have been studied only when species are defined on the nodes of a network. However, in a number of real systems including, e.g., the brain and the climate, dynamical variables are not only defined  on nodes but also on links, faces and  higher-dimensional cells of simplicial or cell complexes, leading to {\em topological signals}.
In this work we study reaction-diffusion processes of topological signals coupled through the  Dirac operator. The Dirac operator allows topological signals of different dimension to interact or cross-diffuse as it projects the topological signals defined on simplices or cells of a given dimension to simplices or cells of one dimension up or one dimension down. By focusing on the framework involving nodes and links we establish the conditions for the emergence of Turing patterns and we show that the latter are never localized only on nodes or only on links of the network. Moreover when the topological signals display Turing pattern their projection does as well.
We validate the theory hereby developed on a benchmark network model and on square lattices with periodic boundary conditions.
\end{abstract}

\maketitle

\section{Introduction}

Nature is a blossoming of patterns, \textcolor{black}{namely spatially heterogeneous structures,} spontaneously emerging from the web of nonlinear interactions existing among the many basic units constituting the system under scrutiny~\cite{PrigogineNicolis1967,pikovsky2001synchronization}. Scholars have developed theories capable to deal with both the case of stationary patterns~\cite{Turing,NM2010,pastorsatorrasvespignani2010} and time varying ones~\cite{kuramoto1975,strogatz2000kuramoto,arenas2008synchronization,Boccaletti,CARLETTI2022112180}. Such research has been developed in the framework of network science~\cite{barabasibook,newmanbook,latora_nicosia_russo_2017,boccaletti2006complex} relying on the assumption that system interactions can be sufficiently well described by using a pairwise representation: the basic units composing the system exhibit their own dynamics, i.e., a local evolution law associated to each node of the network, and then they interact by diffusing or via non-local (long-range) interactions, by using the available links.

Networks however only capture pairwise interactions while higher-order interactions \cite{bianconi2021higher,battiston2020networks,torres2021and,bick2021higher,giusti2016two,salnikov2018simplicial,otter2017roadmap,battiston2022higher} are crucial to describe several empirical systems in physics, biology, neuroscience or social sciences.  Interestingly, recent research taking into  account higher-order interactions is rapidly changing our understanding of the relation between structure and function of complex systems \cite{bianconi2021higher,natphys,majhi2022dynamics}.

Simplicial complexes are higher-order networks that come with extremely rich and useful structures inherited from discrete topology~\cite{bianconi2021higher,nakahara2003geometry,Lim2020}. 
Roughly speaking, a simplicial complex is a topological structure that, besides nodes and links, also contains triangles, i.e., three-body interactions, tetrahedra, i.e., four-body interactions, and so on. \textcolor{black}{Even more generally cell complexes \cite{mulder2018network} also include the other convex polytopes, i.e., not only triangles and tetrahedra but also squares, pentagons, etc. and  hypercubes, orthoplex etc.} 
One can thus consider topological signals defined on nodes and links, but also on higher-order structures~\cite{bianconi2021higher}. 
Examples of topological signals  occur \textcolor{black}{ for instance in neuronal networks, where the interaction between two neurons is mediated by the synaptic signal~\cite{linne2022neuron}.}
\textcolor{black}{Recent scientific literature points out the relevance of edge signals also in large scale brain networks} \cite{faskowitz2022edges,santoro2022unveiling}, and in  biological transportation networks \cite{katifori2010damage,rocks2021hidden}. \textcolor{black}{Edge signals occur also in in power-grids} \cite{witthaut2022collective} or in traffic on a road network \cite{barbarossa2020topological,sardellitti2022topological,schaub2018flow,schaub2021signal}. \
Moreover edge signals might also represent a number of climate data such as currents in the ocean and velocity of wind that can be projected on a suitable triangulation of the Earth surface \cite{schaub2018flow,schaub2021signal}. 
Topological signals can undergo higher-order simplicial  synchronization~\cite{millan2020explosive,carletti2022global,millan2022geometry,torres2020simplicial,ghorbanchian2021higher,calmon2021topological,calmon2022local,petri_hodgeg,deville2020consensus}, and higher-order diffusion \cite{torres2020simplicial,reitz2020higher,ziegler2022balanced}. Moreover datasets of topological signals can be treated with topological signal  processing \cite{barbarossa2020topological,schaub2020random,schaub2021signal} and with topological machine learning tools~\cite{bodnar2021weisfeiler,ebli2020simplicial,roddenberry2019hodgenet,hajij2020cell}. Note that this increasing interest in topological signals occurs while the entire field of dynamical processes on simplicial complexes and hypergraphs is bursting with significant research activity \cite{skardal2019abrupt,skardal2020higher,gambuzza2021stability,kovalenko2021contrarians,alvarez2021evolutionary,lee2021homological,carletti2020random,lucas2020multiorder,tang2022optimizing,zhang2021unified,chutani2021hysteresis,mulas_msf}.

Topological signals of a given dimension can be coupled by the higher-order Laplacians also called Hodge-Laplacians \textcolor{black}{or combinatorial Laplacians}~\cite{bianconi2021higher,josthorak2013spectra,lim2020hodge}.
However the Dirac operator ~\cite{bianconi2021topological,lloyd2016quantum,ameneyro2022quantum,post2009first} is necessary to couple topological signals of different dimension such as interacting signals defined on nodes and links of a network. \textcolor{black}{For instance the dynamics of neuronal networks can be modeled by using two different topological signals: one defined on the nodes (the activity of each neuron) and the other defined on the edges (the neurotransmitter current across each synapse).}
Interestingly, the  Dirac synchronization which stems from the adoption of the Dirac operator to couple topological signals of different dimension, provides a topological and local pathway towards explosive synchronization and rhythmic phases~\cite{calmon2021topological,calmon2022local}.

In this paper we propose a framework to reveal Turing patterns of reacting species described by topological signals defined on the \textcolor{black}{cells} of different dimensions (nodes, links, triangles, squares) coupled through the Dirac operator. Our main goal is to consider reaction-diffusion systems~\cite{Murray2001} and extend the Turing theory developed so far on networked systems~\cite{NM2010} to the framework of simplicial \textcolor{black}{and cell} complexes.

Turing's original framework involved two reacting species whose stable homogeneous equilibrium can turn out unstable once the species are allowed to diffuse and suitable conditions of the species \textcolor{black}{diffusion coefficients} are assumed~\cite{Turing}. Gierer and Meinhardt later emphasized that for the Turing instability to set up, one of the two species needs to be an activator while the other should be an inhibitor, and moreover the latter needs to diffuse much faster than the former~\cite{GiererMeinhardt}. The theory was successively extended to regular lattices by Othmer and Scriven~\cite{OS1971} and finally to complex networks by Nakao and Mikhailov~\cite{NM2010}. \textcolor{black}{Let us emphasize that network patterns are equilibrium states of the system with a dependence on the node.} The latter framework has been further expanded considering directed networks \cite{Asllani1}, multiplex \cite{Asllani2014}, temporal networks \cite{PABFC2017} and non-normal networks \cite{jtb}, just to mention a few. In all the above settings, the two species react in each node while diffusing through the links. 
For signals defined exclusively on the nodes  cross-diffusion terms have been been introduced in~\cite{fanelli_cianci,busiello_turing}.
Turing patterns on higher-order structures  have been recently studied in~\cite{carletti2020dynamical,muologallo}. Note however that our approach is different because in those works the dynamics is restricted to nodes, while links and high-order structures support the generalized diffusion.

In this paper we provide a general theory describing reaction-diffusion systems of topological signals of different dimension (i.e., defined on nodes, links, triangles, squares, etc.) coupled with the Dirac operator.
In particular, we consider two different settings. \textcolor{black}{In the first case we assume} the reaction term \textcolor{black}{to be} solely responsible for the coupling of signals of different dimension \textcolor{black}{and the diffusion term is modeled by the \textcolor{black}{Hodge-Laplacians}.} \textcolor{black}{In the second case, we assume} the diffusion also \textcolor{black}{to} include cross-diffusion terms coupling the dynamics of signals in different dimension.
For the sake of simplicity, in this work, we will focus our analysis to the case of coupled nodes and links signals which is arguably also the most relevant to applications. \textcolor{black}{Indeed it is a common scenario to have localized reactions and quantities produced in the nodes, to flow across links connecting couples of node; in some cases links themselves are dynamical entities, whose behavior influence the local reactions but can also be in turn influenced by the latter}. We derive the conditions under which stable Turing patterns can be observed and we highlight the differences between the dynamics with and without cross-diffusion terms. The analytical results derived in general are presented with applications to square lattices with periodic boundary conditions and validated by numerical simulations on a benchmark network.

The paper is structured as follows. In Sec II we outline a general theoretical framework for investigating Turing patterns of topological signals, and we  distinguish the case in which there is only a Dirac reaction term while diffusion is dictated by Hodge-Laplacians and the case in which we introduce also Dirac cross-diffusion terms describing diffusion processes among signals defined on different dimensions. In Sec. III and IV we focus on topological signals defined on nodes and links of the network and we define the conditions for the onset of the Turing instability when only a Dirac reaction term is considered (Sec. III) and when additionally Dirac cross-diffusion terms are introduced (Sec.IV). The theoretical insights gained in Sec. III and IV are tested and validated on a benchmark model. Finally, in Sec. V we provide the concluding remarks. The paper is enriched with few appendices providing background information on algebraic topology, some details of the derivations discussed in the main body of the work and simulations results on  Turing patterns of topological signals defined on nodes and links of a square lattice with periodic boundary conditions.\\

\section{Turing theory for topological signals}
We are interested in studying reaction-diffusion systems defined on simplicial \textcolor{black}{and cell} complexes (for an introduction to \textcolor{black}{such topological structures} and their main properties see Appendix~\ref{appA:alg_topology}). This entails defining appropriate reaction and diffusion terms. In a network the reaction term is localized on nodes, where the interacting species can be found. When the interacting species are associated to simplices of different dimension, a Dirac reaction term that uses the Dirac operator is required to allow topological signals of different dimension to interact.
In a network, concentrations can flow from one node to one of its neighbors, passing through links, namely the structure one dimension above. A similar idea can be thought in simplicial complexes: quantities defined on links can flow among links by using the faces they share, hence again the structures one dimension above. There is however a second possibility: they can use structures one dimension below, i.e., nodes, to communicate. Such processes can be described by introducing the \textcolor{black}{Hodge-Laplacian operator} which describes uncoupled diffusion of topological signals of any given dimension. However Hodge-Laplacians describe diffusion terms that act on topological signals of any given dimension separately.
Requiring a diffusive coupling of topological signals of different dimension can be only achieved by considering Dirac cross-diffusion terms which involve odd powers of the Dirac operator. Specifically, this includes cross-diffusion terms that are linear or cubic in the Dirac operator.
\begin{figure*}[!htb]
\centering
		\includegraphics[width = 1\textwidth, angle=0]{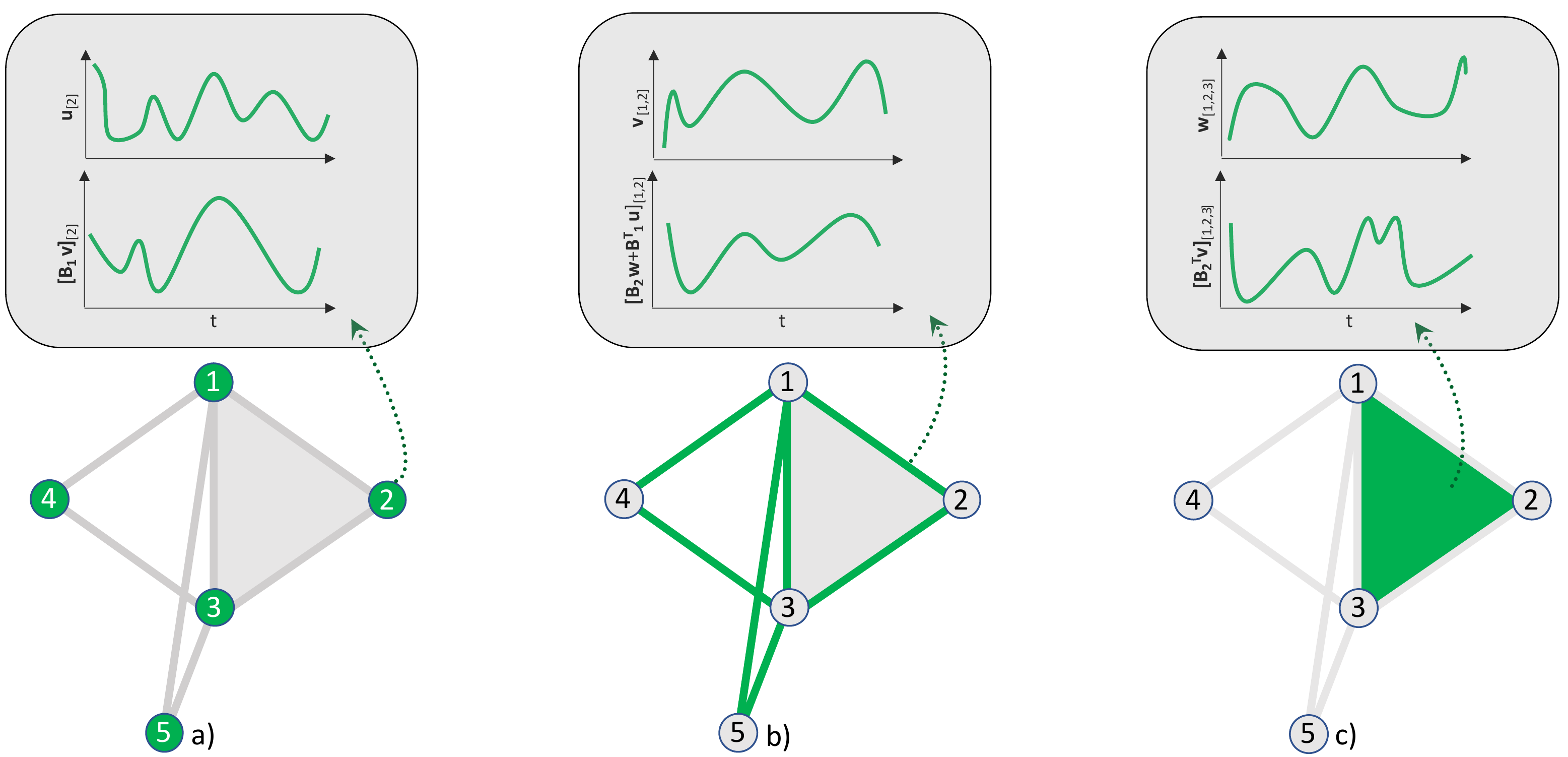}
	\caption{We schematically represent the dynamical state of a simplicial complex encoded by the vector  $\bm\Phi=(u,b,w)^{\top}$  and the vector  $\bm\Psi=\mathcal{D}\bm\Phi=(\hat{u},\hat{v},\hat{w})^{\top}$. In particular we represent topological signals and projected topological signals supported on $0$, $1$ and $2$-simplices respectively in panels a), b), and c). The Dirac operator $\mathcal{D}$ projects the topological signals of each dimension either one dimension up or one dimension down, and leads to projected components defined on nodes $(\hat{u}={\bf B}_1v$, links $\hat{v}={\bf B}_2w+{\bf B}_1^{\top}v$, and triangles $\hat{w}={\bf B}_2^{\top}v$. Here $\hat{u}={\bf B}_1v$ describes the link signals  projected on the nodes; ${\bf B}_1^{\top}u$ indicates  the  irrotational component of $\hat{v}$ and describes the projection of the node signals on the links; ${\bf B}_2w$ indicates the solenoidal component of $\hat{v}$ and describes the projection of the triangle signals on the links; finally   ${\bf B}_2^{\top}v$ describes the projection of the link signals on the  triangles.}
	\label{f:fig1}
\end{figure*}

Here we propose a \textcolor{black}{theory of} Turing \textcolor{black}{instability} for topological signals and to this end we consider a simplicial \textcolor{black}{and cell} complexes of dimension $d$ and  species living on nodes, links, triangles, etc.
In the present terminology, the concentration of the  species living on nodes is a $0$-topological signal while the concentration of the species defined on links is a $1$-topological signal etc.
The dynamical state of the \textcolor{black}{structures we are considering} is described by a vector $\Phi$ which is the direct sum of all topological signals defined on the simplicial  \textcolor{black}{or cell} complex.
For example in a $d=2$ dimensional \textcolor{black}{cell} complex with $N_0$ nodes, $N_1$ links and $N_2$ \textcolor{black}{$2$-dimensional cells (such as triangles, squares, pentagons, etc.)} we have 
\begin{equation}
\Phi=\begin{pmatrix}u\\v\\w\end{pmatrix},\label{Phi}
\end{equation}
where $u\in \mathbb{R}^{N_0}, v\in \mathbb{R}^{N_1}, w\in \mathbb{R}^{N_2}$ are the vectors of concentration of species defined on nodes, links and \textcolor{black}{$2$-dimensional cells} respectively.
These signals can only interact with each other when we consider their projection to simplices of one dimension up or one dimension down. This projection is performed by applying the Dirac operator $\mathcal{D}$ to $\Phi$ obtaining new (projected) signals (for the definition of the Dirac operator see Appendix \ref{appA:alg_topology}), i.e.,
\begin{equation}\label{def_Psi}
\Psi=\mathcal{D}\Phi=\begin{pmatrix}\hat{u}\\ \hat{v}\\\hat{w}\end{pmatrix},\end{equation} where 
$\hat{u}\in \mathbb{R}^{N_0}, \hat{v}\in \mathbb{R}^{N_1}, \hat{w}\in \mathbb{R}^{N_2}$ are defined on nodes, links and \textcolor{black}{$2$-dimensional cells} respectively.
In a \textcolor{black}{general cell}  complex of dimension $d=2$ the Dirac operator $\mathcal{D}$ is a $M\times M$ matrix with $M=N_0+N_1+N_2$ which can be expressed in terms of the incidence matrices ${\bf B}_1,{\bf B}_2$ (defined in Appendix \ref{appA:alg_topology}) and their transpose as 
\begin{equation}
\mathcal{D}=\begin{pmatrix}0& {\bf B}_1&0\\
{\bf B}_1^{\top}&0&{\bf B}_2\\
0&{\bf B}_2^{\top}&0 \end{pmatrix}. 
\end{equation}
We therefore obtain that the projected signal $\Psi$ is given by 
\begin{equation}\label{def_Psi2}
\Psi=\mathcal{D}\Phi=\begin{pmatrix}\hat{u}\\ \hat{v}\\\hat{w}\end{pmatrix}=\begin{pmatrix}{\bf B}_1v\\{\bf B}_1^{\top}u+{\bf B}_2w\\{\bf B}_2^{\top}v\end{pmatrix},
\end{equation}
where ${\bf B}_1^{\top}u$  and  ${\bf B}_2w$ describe the irrotational part  and the solenoidal part of the  link signal $\hat{v}$.
Therefore, the dynamical state of the \textcolor{black}{cell} complex comprises both the topological signals $\Phi$ and their projections $\Psi=\mathcal{D}\Phi$ (see Figure \ref{f:fig1} for a schematic illustration).
Note that the Dirac operator can be seen as the ``square root" of the higher-order or Hodge-Laplacian operator $\mathcal{L}$ as 
\begin{equation}
\mathcal{L}=\mathcal{D}^2=\begin{pmatrix}{\bf L}_0&0&0\\
0&{\bf L}_1&0\\
0&0&{\bf L}_2 \end{pmatrix},
\end{equation}
where ${\bf L}_{0}={\bf B}_1{\bf B}_1^{\top},{\bf L}_{1}={\bf B}_1^{\top}{\bf B}_1+{\bf B}_2{\bf B}_2^{\top}$ and ${\bf L}_2={\bf B}_2^{\top}{\bf B}_2$ are the \textcolor{black}{Hodge-}Laplacians acting on topological signals of dimension zero, one, and two respectively and describing higher-order diffusion  (for details see Appendix \ref{appA:alg_topology}) \cite{torres2020simplicial,reitz2020higher,ziegler2022balanced}. In particular, \textcolor{black}{in the case of a simplicial complex} we have that ${\bf L}_0$ describes diffusion from nodes to nodes through links, ${\bf L}_1$ describes diffusion from links to links either through nodes or through triangles and,  ${\bf L}_2$ describes diffusion from triangles to triangles through links.
Here we propose a Turing \textcolor{black}{instability} theory for topological signals where the topological signals $\bm\Phi$ can be coupled to the projected topological signals $\Psi$ either through a Dirac reaction term or through a Dirac diffusion term or both.
In presence of a Dirac reaction term and a Laplacian diffusion term, the reaction-diffusion process of topological signal is defined as 
\begin{equation}\label{Dirac_reaction}
\dot{\Phi}=F(\Phi,\mathcal{D}\Phi)-\gamma\mathcal{L}\Phi,
\end{equation}
where  $F(\Phi,\mathcal{D}\Phi)$ is the Dirac reaction term coupling each topological signal of dimension $n$ with the nearby topological signals of  dimension $n+1$ or $n-1$ projected to dimension $n$.
In particular  $F(\Phi,\mathcal{D}\Phi)$ here indicates a generic nonlinear function, assumed to be applied component-wise on the vectors. For instance for $d=2$ we have 
\begin{equation}
F(\Phi,\mathcal{D}\Phi)=\left(\begin{matrix} 
			f_0(u, \mathbf{B}_1 v) \\
			f_1(v,\mathbf{B}_1^{\top} u+\mathbf {B}_2w)\\
		    f_2(w,\mathbf{B}_2^{\top}v).
			\end{matrix}\right),
\end{equation}
where  $f_{n}(x,y)$ are   nonlinear functions, such that  $f_1(u, \mathbf{B}_1v)=(f_1(u_1, (\mathbf{B}_1v)_1),\dots,f_1(u_{N_0}, (\mathbf{B}_1v))_{N_0})$ etc.
The matrix $\gamma$ in Eq.(\ref{Dirac_reaction}) is a diagonal matrix  
\begin{equation}
\gamma=\begin{pmatrix}D_0&0&0\\
0&D_1&0\\0&0&D_2\end{pmatrix},
\end{equation}
where $D_n$ is the diffusion constant acting on topological signals of order $n$.
Therefore Eq.(\ref{Dirac_reaction}) describes topological signals defined on the \textcolor{black}{cells} of the \textcolor{black}{cell} complex that react with the projection of the topological signals defined in different dimension while undergoing higher-order diffusion.

Note that from the dynamical system given by Eq.(\ref{Dirac_reaction}) one can derive the dynamics of the projected signal 
 $\Psi=\mathcal{D}\Phi$ which is given by 
\begin{equation}
\dot{\Psi}=\hat{F}(\Phi,\Psi)-\mathcal{D}\gamma\mathcal{D}\Psi,
\end{equation}
where $\hat{F}(\Phi,\Psi)=\mathcal{D}F(\Phi,\Psi)$.
In the case of diffusion constants independent on the order of the simplices, i.e., for $D_k=D$, this equation reduces to  
\begin{equation}
\dot{\Psi}=\hat{F}(\Phi,\Psi)-\gamma\mathcal{L}\Psi.\end{equation}
Therefore in this case the dynamics of the projected signal is the same as the dynamics of the signal $\Phi$ (Eq. (\ref{Dirac_reaction})) provided that $F(\Psi,\Phi)=\hat{F}(\Phi,\Psi)=\mathcal{D}F(\Phi,\Psi)$ as for instance in the case of square lattices with periodic boundary conditions.

We now consider Dirac cross-diffusion terms enforcing diffusion of signals across different dimensions.

In particular, we consider including a linear or a cubic Dirac cross-diffusion term which are proportional to a linear or cubic power of the Dirac operator. \textcolor{black}{Let us observe that this is a natural choice, since as already observed, the second power of the Dirac operator is \textcolor{black}{a diagonal matrix containing Hodge-Laplacians on its diagonal}.}
In the case of a linear Dirac cross-diffusion term, the reaction-diffusion dynamics takes the form
\begin{equation}\label{Dirac_diffusion_linear}
\dot{\Phi}=F(\Phi,\mathcal{D}\Phi)-\tilde{\gamma}\mathcal{D}\Phi-{\gamma}\mathcal{L}\Phi,
\end{equation} where $\tilde{\gamma}$ is the diagonal matrix of cross-diffusion constants $\tilde{D}_n$,
\begin{equation}
 \tilde{\gamma}=\begin{pmatrix}\tilde{D}_0&0&0\\
0&\tilde{D}_1&0\\0&0&\tilde{D}_2\end{pmatrix}.
\end{equation}
In this case, the corresponding projected signals $\Psi=\mathcal{D}\Phi$ obey the dynamical system of equations
\begin{equation}
\dot{\Psi}=\hat{F}(\Phi,\Psi)-\mathcal{D}\tilde{\gamma}\Psi-\mathcal{D}{\gamma}\mathcal{D}\Psi.
\end{equation}
If the diffusion and cross-diffusion constants are the same and $\gamma$ and $\tilde{\gamma}$ are proportional to the identity matrix, then we have that both $\gamma $ and $\tilde{\gamma}$ commute with the Dirac operator $\mathcal{D}$ and the dynamics of projected signals becomes
\begin{equation}
\dot{\Psi}=\hat{F}(\Phi,\Psi)-\tilde{\gamma}\mathcal{D}\Psi-{\gamma}\mathcal{L}\Psi.
\end{equation}
Therefore, in this case too, as long as  $\hat{F}(\Phi,\psi)=\mathcal{D}F(\Phi,\Psi)$ can be written as  the reaction term $F(\Phi,\Psi)$ (as it happens for square lattices with periodic boundary conditions for example) the equation for the signal is equal to the equation for the projected signals. 
In the case of a cubic Dirac cross-diffusion term, we have instead that
\begin{equation}\label{Dirac_diffusion_cubic}
\dot{\Phi}=F(\Phi,\mathcal{D}\Phi)-\gamma\mathcal{L}\Phi-\tilde{\gamma}\mathcal{D}^3\Phi.
\end{equation}
The corresponding projected dynamics reads,
\begin{equation}
\dot{\Psi}=\hat{F}(\Phi,\Psi)-\mathcal{D}{\gamma}\mathcal{D}\Psi-\mathcal{D}\tilde{\gamma}\mathcal{L}\Psi,
\end{equation}
which reduces to 
\begin{equation}
\dot{\Psi}=\hat{F}(\Phi,\Psi)-{\gamma}\mathcal{L}\Psi-\tilde{\gamma}\mathcal{D}^3\Psi,
\end{equation}
when\textcolor{black}{, again,} both $\gamma$ and $\tilde{\gamma}$ are proportional to the identity matrix.

In all the considered cases, the Turing mechanism requires the presence of a stable homogeneous equilibrium once the diffusion part is silenced. Such state turns out unstable for suitable values of the diffusion coefficients and conditions on the underlying topology. \textcolor{black}{Eventually, arbitrarily small initial perturbations around the homogeneous state will exponentially grow and ultimately return a pattern, i.e., a spatially heterogeneous solution.}

When dealing with topological signals, a necessary condition is that the homogeneous state vector $h=(1,\dots,1)^\top$ is in the kernel of the Dirac operator  $h\in \mbox{ker}(\mathcal{D})$ or, equivalently,
\begin{equation}\label{kernel1}
\mathcal{D}h=0. 
\end{equation}
In  conventional node to node diffusion case, in which only the node signal is considered, such condition is always satisfied for a connected network. However when the state vector includes both nodes and links signals Eq.\eqref{kernel1}  accounts to require
\begin{equation}
\label{eq:kerprop}
    \mathbf{B}_1 \hat{h}=0\text{ and }\mathbf{B}_2^\top \hat{h}=0\,
\end{equation}
where $\hat{h}=(1,1\ldots, 1)^{\top}$ is a homogeneous  $N_1$-dimensional column vector  defined on the links of the network.

By assuming to have a $1$-simplicial complex, (i.e., a network) we discard the presence of \textcolor{black}{$2$-dimensional cells (such as triangles, squares, pentagons,etc.)}. In that case $\mathbf{B}_2=0$, and the second of the conditions in Eq.\eqref{eq:kerprop} is trivially satisfied. Let us now focus on the remaining condition. Tackling this problem becomes much easier by noticing that the $i$-th row of the boundary operator is equal to minus the divergence of node $i$. Such equivalence, proved in \cite{Lim2020}, can be exploited to construct a simplicial complex with the wanted property.

By requiring that every node has an equal amount of in-coming and out-going links, we thus ensure that a homogeneous signal, namely an edge-flow directed as indicated by the links orientation~\footnote{\textcolor{black}{Let us stress that we are dealing with undirected network and thus the incoming / outgoing edges are defined with respect to the ordering of the simplicial or cell complex.}}, has zero divergence. To sum up, the following analysis grounded on the conditions given in Eq. \eqref{eq:kerprop}, holds for every \textcolor{black}{network ($1$-dimensional cell complex)}  whose nodes have an even number of connected edges. Notably examples of these networks are square lattices with periodic boundary conditions.

\textcolor{black}{Note that the analogous condition applying to   $2$-dimensional cell complexes is much more demanding. In particular no $2$-dimensional simplicial complex admits an homogeneous eigenvector in the kernel of the Dirac operator. However it was recently shown \cite{carletti2022global} that $2$-dimensional cell complexes built from square lattices with periodic boundary conditions obey this property. More generally it is possible to show that $d$-dimensional cell complexes built from $d$-dimensional square lattices obey this property for any dimension $d$.}
\section{Interacting topological signals of nodes and links with Dirac reaction term}\label{sec:diff_coupling}
\subsection{Conditions for the onset of the Turing instability}
In this section we focus on reaction-diffusion systems involving  topological signals defined on the nodes and on the links of a network.
Our goal is to derive the dispersion relation\textcolor{black}{, roughly speaking the largest Lyapunov exponent of the homogeneous state considered as a function of the model parameters and of the topological structure. This allows} us to determine the conditions for the Turing instability onset in the presence exclusively of a Dirac reaction term that couples the two topological signals of different dimension, \textcolor{black}{while the diffusion part is modeled \textcolor{black}{with the relevant Hodge-Laplacians} }, i.e., driven by Eq.(\ref{Dirac_reaction}) which we rewrite here for convenience
\begin{equation}\label{Dirac_reaction2}
\dot{\Phi}=F(\Phi,\mathcal{D}\Phi)-\gamma\mathcal{L}\Phi.
\end{equation}

In a network we have $\Phi=(u,v)^{\top}$ and ${F}(\Phi,\mathcal{D}\Phi)=\left({f}(u, \mathbf{B}_1 v),{g}(v,\mathbf{B}_1^{\top} u)\right)^{\top}$ where $f$ and $g$ are two generic nonlinear functions, assumed to be applied component-wise on the vectors, i.e., $f(u, \mathbf{B}_1v)=(f(u_1, (\mathbf{B}_1v)_1),\dots,f(u_{N_0}, (\mathbf{B}_1v))_{N_0})$. 
Here $\gamma$ reduces to the $(N_0+N_1)\times (N_0+N_1)$ block diagonal matrix with structure
\begin{equation}
\gamma = \begin{pmatrix}
D_0 {\bf I}_{N_0} & 0 \\
0 & D_1 {\bf I}_{N_1}
\end{pmatrix},
\end{equation}
where $D_0$ and $D_1$ indicate the diffusion constants of the species defined on nodes and links respectively \textcolor{black}{and ${\bf I}_{N_a}$ indicates the $N_a\times N_a$ identity matrix, $a=0,1$}.
The  Dirac operator $\mathcal{D}$ and the \textcolor{black}{Hodge-}Laplacian operator $\mathcal{L}$ are defined as the $(N_0+N_1)\times (N_0+N_1)$ matrices with block structure
\begin{equation}\mathcal{D}=\begin{pmatrix} 0 & \mathbf{B}_1 \\ \mathbf{B}_1^\top & 0 \end{pmatrix},\quad \mathcal{L}=\mathcal{D}^2=\begin{pmatrix}
    \mathbf{L}_0 & 0 \\
    0 & \mathbf{L}_1
\end{pmatrix}.
\end{equation}
If follows that the  dynamics  driven by Eq.(\ref{Dirac_reaction2}) can be rewritten explicitly as
\begin{equation}
\begin{aligned}
&\frac{du}{dt}=f(u, \mathbf{B}_1v)-D_0\lap_{0} u, \\
&\frac{dv}{dt}=g\left(v,\mathbf{B}_1^{\top} u\right) - D_1\lap_{1} v \,
\end{aligned}\label{eq:syst1}
\end{equation} 
where $D_0>0$ (resp. $D_1>0$) is the diffusive coefficient of species $u$ (resp. $v$). \textcolor{black}{For instance, resuming the biological example from the introduction where neurotransmitters concentration and neuronal activity are schematized by topological signals,
we can think of $v$ as the synaptic signal, and of $u$ as a neuron signal. In this setting the Dirac operator is capable of properly connect the lower and higher dimensional signals, by acting as an effective and simple dynamical operator.}
In the spirit of Turing theory, let us silence the diffusive terms and look for a homogeneous solutions, i.e., the existence of $u^*=u_0h$ and $v^*=v_0h$, for some constants $u_0$ and $v_0$. Because of the assumption on the underlying simplex, we have $\mathbf{B}_1v^*=0 $ and $\mathbf{B}_1^\top u^*=0$. The existence of a homogeneous fixed point reverberates on the structure of $f,g$ such that
\begin{equation}
0=f(u^*, 0) \text{ and } 0=g( v^*,0)\, ,
\end{equation}
which in turn yields that $u_0$ and $v_0$ are solutions of $f(u_0, 0)=g(v_0,0)=0$.

To study the stability feature of the homogeneous equilibrium, we consider a homogeneous perturbation about the latter, $\delta u = u-u^*$ and $\delta v = v-v^*$. Hence by linearizing~\eqref{eq:syst1} we obtain
\begin{equation}	
\begin{aligned}
	&\frac{d\delta u}{dt}=\partial_{u} f(u^*,0) \delta u, \\
	&\frac{d\delta v}{dt}=\partial_v g(v^*,0) \delta v\, ,
\end{aligned}
\end{equation}
where we used again the conditions $h=(1,\dots,1)^\top\in \ker\mathbf{B}_1$ and $h\in \ker\mathbf{B}^\top_1$ to remove some terms in the previous equation. The condition for the stability is thus
\begin{equation}
\label{eq:stabilityhom}
\partial_{u} f(u^*,0) <0\text{ and }\partial_{v} g(v^*,0)<0\, .
\end{equation}

Let us observe that Eq.~\eqref{eq:stabilityhom} implies that both species are self inhibitors, this is the result of the peculiar form of Eq.~\eqref{eq:syst1}, and of the assumption $\mathbf{B}_1v^*=0 $ and $\mathbf{B}_1^\top u^*=0$ which ultimately decouples the dynamics of the two species in the linear regime. This is at odd with the classical Turing instability where patterns can never emerge in the  inhibitor-inhibitor setting, unless some additional assumptions are made \cite{jop_carletti}.

We now focus on the stability of such equilibrium once subjected to heterogeneous perturbations, hence not in the kernels of $\lap_0$ and $\lap_1$.
Let us linearize Eq.~\eqref{eq:syst1} about the equilibrium solution, \textcolor{black}{by} obtaining
\begin{equation}\label{linearised}	
\begin{aligned}
	&\frac{d\delta u}{dt}=\left(\partial_{u} f\right) \delta u+\left(\partial_{\mathbf{B}_1v} f\right) \mathbf{B}_1 \delta v-D_0\lap_{0}\delta u, \\
	&\frac{d\delta v}{dt}=(\partial_{\mathbf{B}_1^\top u}g) \mathbf{B}_1^\top \delta u+(\partial_v g) \delta v-D_1\lap_{1}\delta v \, ,
\end{aligned}
\end{equation}
where $\partial_{\mathbf{B}_1v}f$ and $\partial_{\mathbf{B}_1v}f$ denote the scalars indicating the derivative of  $f$, $g$ with respect to their second argument, (i.e., the projected higher and lower dimensional signal respectively) calculated at the homogeneous stationary solution.

We now note that the network Laplacians $\lap_0={\bf B}_1{\bf B}_1^{\top}$ and $\lap_1={\bf B}_1^{\top}{\bf B}_1$ are isospectral, i.e., they have the same non-zero spectrum. The $\hat{N}$ non-zero eigenvalues $\Lambda_0^k$ with $1\leq k\leq \hat{N}$ of $\lap_0$ and $\lap_1$ can be expressed as the square of the singular values $b_k$ of ${\bf B}_1$, i.e., $\Lambda_0^k=b_k^2.$ The eigenvectors $\psi_0^m$ and $\psi_1^m$ of $\lap_0$ and $\lap_1$  can be adopted as a basis to perform the singular value decomposition of ${\bf B}_1$. On a connected network these eigenvectors include the eigenvectors $\psi_0^k$ and $\psi_1^k$ corresponding to the non-zero eigenvalue $\Lambda_0^k=\Lambda_1^k=b_k^2$, the eigenvector $\phi_0^h=(1,\ldots,1)^{\top}$ of $\lap_0$ associated to the zero  eigenvalue $\Lambda_0=0$ and the eigenvectors $\psi_1^{l}$ associated the zero eigenvalues $\Lambda_1^l=0$ of $\lap_1$. Interestingly the eigenvectors $\psi_0^k$ and $\psi_1^k$ associated to the eigenvalue $\Lambda_0^k=\Lambda_1^k=b_k^2>0$ obey
\begin{equation}
    {\bf B}_1\psi_1^k=b_k\psi_0^k, \quad {\bf B}_1^{\top}\psi_0^k=b_k\psi_1^k. 
\end{equation}

Using these results, the signals $\delta u$ and $\delta v$, as well as the projected signals $\delta\hat{u}={\bf B}_1\delta v$ and $\delta\hat{v}={\bf B}_1^{\top}\delta u$, can be projected onto the basis of  the eigenvectors $\psi_n^m$ of $\lap_n$ (with $n=0,1$ for the analyzed case) corresponding to the non-zero eigenvalues $\Lambda_0^k=b_k^2$. We obtain
\begin{eqnarray}\label{deltas}
	\langle \psi_0^k,\delta u\rangle=\delta \hat{u}_{k} , &\quad& \langle \psi_1^k,\delta v\rangle=\delta\hat{v}_{k} \, ,\\
		\langle \psi_0^k,{\bf B}_1\delta {v}\rangle= b_k\delta\hat{v}_{k} , &\quad &\langle \psi_1^k,{\bf B}_1^{\top}\delta u\rangle= b_k\delta\hat{u}_{k} ,\label{deltas2}
\end{eqnarray}
where $\langle\cdot,\cdot\rangle$ denotes the scalar product.
By using Eq.\eqref{deltas} and Eq.\eqref{deltas2}, we can project in Eq.(\ref{linearised}) the equations for $\delta u$ onto $ {\psi^{k}_0}^\top $ and the ones for $\delta v$ on ${\psi^{k}_1}^\top$, with $k$ such that $ \Lambda_0^{k}=\Lambda_1^{k}=b_k^2\neq0$, to eventually obtain:
\begin{equation}\label{linear_diffusion}
	\begin{aligned}
			\frac{d\delta\hat{ u}_{k}}{dt}&=\left(\partial_{u} f\right) \delta\hat{ u}_{k} +\left(\partial_{\mathbf{B}_1v} f\right) b_k \delta\hat{v}_{k}- D_0b_k^2 \delta\hat{ u}_{k},\\
			\frac{d\delta\hat{ v}_{k}}{dt}&=\left(\partial_{v} g\right) \delta\hat{ v}_{k} +\left(\partial_{\mathbf{B}_1^\top u} g\right) b_k \delta\hat{ u}_{k} - D_1 b_k^2 \delta\hat{ v}_{k}\, .
	\end{aligned}
\end{equation}

It is interesting to notice that the leftover modes are those associated to the eigenvectors spanning the kernel space of both $\lap_{0}$ and $\lap_{1}$. Since in the relevant case of a connected network, the eigenvector associated to the zero eigenvalue is the homogeneous one, i.e it is aligned to the stationary state $u^*$ of the nodes, it follows that $ \delta u$ will never have a component along this eigenvector. However, we need to consider the projection of $\delta v$ onto the eigenvectors $\psi_1^l$ associated to the zero eigenvalues of $\lap_1$, \textcolor{black}{by} obtaining
\begin{equation}\label{always_stable}
	\frac{d\delta\hat{ v}_{l}}{dt}=\left(\partial_{v} g\right)\delta \hat{ v}_{l}\, .
\end{equation}
Hence these modes are always stable due to the second condition in Eq.~\eqref{eq:stabilityhom}.

The instability is realized if the linear system~\eqref{linear_diffusion} admits at least one unstable mode; more precisely we have to compute the eigenvalues of the matrix
\begin{equation}
	\mathbf{J}_k=\left(\begin{matrix} 
		\partial_{u} f- D_0b_k^2  &b_k \partial_{\mathbf{B}_1v} f\\
		b_k\partial_{\mathbf{B}_1^\top u} g &  \partial_{v} g- D_1 b_k^2 
	\end{matrix}\right)\, ,
\end{equation}
and determine if there is $k$ for which the associated eigenvalue, $\lambda(b_k)$, has a positive real part. Let us notice that the latter is usually named dispersion relation in the literature. The eigenvalues of $\mathbf{J}_k$ can be obtained by solving
\begin{equation}\label{eq:lambda}
	\lambda^{2}+\lambda \Gamma_{1}\left(b_k^{2}\right)+\Gamma_{2}\left(b_k^{2}\right)=0\, .
\end{equation}
where  $\Gamma_1\left(b_k^{2}\right)$ and $\Gamma_{2}\left(b_k^{2}\right)$ are given by 
\begin{eqnarray}
\Gamma_{1}\left(b_k^{2}\right)&=&b_k^2(D_1+D_0)- (\partial_v g+\partial_u f),\label{eq:term0}\\
\Gamma_{2}\left(b_k^{2}\right)&=&a_2b_k^4+a_1b_k^2+a_0,\label{eq:term}
\end{eqnarray}
with 
\begin{equation}
\begin{aligned}
&a_2=D_0 D_1, \\
&a_1=-\left(D_1\partial_{u} f+ D_0\partial_v g+\partial_{\mathbf{B}_1^\top u} g\ \partial_{\mathbf{B}_1v} f\right), \\
&a_0={\partial_{u} f \  \partial_v g}.
\end{aligned}\label{a}
\end{equation}

Since both the leading coefficient of Eq.\eqref{eq:lambda} and $\Gamma_1(b_k^2)$ are positive, the existence of a solution with positive real part requires that $\Gamma_2(b_k^2)<0$ for some $k$. Let us observe that $\Gamma_2(b_k^2)$ given by Eq. (\ref{eq:term}) is a parabola in $b_k^2$
with positive concavity, $a_2=D_0D_1>0$, and positive constant term, $a_0=\partial_{u} f \  \partial_v g>0$. Therefore, to  satisfy the condition $\Gamma_2(b_k)<0$ with a real $b_k$, a necessary condition is
\begin{equation}\label{turing2}
	D_0\partial_v g+D_1\partial_{u} f+\partial_{\mathbf{B}_1^\top u} g\ \partial_{\mathbf{B}_1v} f>0.
\end{equation}

By using these conditions we can guarantee that $\Gamma_2(b_k^2)<0$ if the minimum of the parabola is negative. A straightforward computation returns the condition  
\begin{equation}\label{turing3}
 {\left(D_0\partial_v g+D_1\partial_{u} f+\partial_{\mathbf{B}_1^\top u} g \partial_{\mathbf{B}_1v} f \right)^2}>{4D_0D_1}\partial_{u} f \partial_v g \, .
\end{equation}
Let us observe that differently from the classical Turing framework, such condition depends on the diffusive coefficients separately and not on their ratio.

In conclusion, we have hence found the conditions for the onset of Turing instability for topological signals whose dynamics is described by Eq.~\eqref{eq:syst1}, namely the stability of the homogeneous solution given by Eq.~\eqref{eq:stabilityhom} and the existence of at least one unstable mode according to Eqs.~\eqref{turing2} and~\eqref{turing3}. Moreover the roots of Eq.~\eqref{eq:lambda}
are given by $\lambda_{1,2}=-\Gamma_1 \pm \sqrt{\Gamma_1^2-4\Gamma_2}$, but $\Gamma_1>0$ and $\Gamma_2<0$, and thus $\lambda_{1,2}$ are real numbers. Consequently, the corresponding  patterns are stationary.

Let us now note that as expected, when the topological signals on nodes and links are not coupled by the Dirac reaction term, i.e., when 
\begin{equation}
F(\Phi,\mathcal{D}\Phi)=F(\Phi)=\begin{pmatrix}f(u)\\g(v)\end{pmatrix},
\end{equation}
we can never have Turing patterns. In fact in this case we would have $\partial_{{\bf B}_1^\top u} g=0, \partial_{{\bf B}_1v}f=0$ and Eq.~\eqref{turing2} cannot be satisfied together with Eq.~\eqref{eq:stabilityhom}.
A major result of this study is that the Turing instability of the topological signals of a network will be never localized only on nodes or only on links but will always involve both nodes and links signals. Moreover, we also obtain that if the original signals $\Phi=(u,v)^{\top}$ display a Turing pattern, the projected dynamics of $\mathcal{D}\Phi=({\bf B}_1v,{\bf B}_1^{\top}u)^\top$ also does. 

\subsection{Numerical results on a benchmark network}
The aim of this section is to validate the above results with a numerical study. To focus on the novelty of the framework and to remove unnecessary complicated features, we will build a toy model with cubic nonlinearities to test our theory (see Appendix \ref{appB:square_lattice} for additional results on topological Turing patterns on the square lattice with periodic boundary conditions). By keeping the same notation as before, i.e., $u$ is the signal on the nodes and $v$ that on the links, the equations of our model read
\begin{eqnarray}
\dot{u}=-a u-bu^3+c \mathbf{B}_1 v - D_0\lap_0 u, \nonumber \\ 
 \dot{v}= -\alpha v-\beta v^3 +\gamma \mathbf{B}_1^{\top}u - D_1\lap_1 v,\label{eq:model}
\end{eqnarray} where $a,b,c,\alpha,\beta,\gamma$ are non-negative real parameters. 

System~\eqref{eq:model} admits $(u_0,v_0)=(0,0)$ as equilibrium point. By computing the Jacobian of the system evaluated at this point, we get
\begin{equation*}
\mathbf{J}_0=\left( \begin{matrix}
\partial_{u} f & \partial_{\mathbf{B}_1v} f\\ \partial_{\mathbf{B}_1^\top u} g & \partial_v g\end{matrix}\right)=\left( \begin{matrix}
-a  &c\\ \gamma & -\alpha\end{matrix}\right)\, .
\end{equation*}
The system exhibits a Turing instability if the above parameters satisfy the conditions~\eqref{eq:stabilityhom},~\eqref{turing2} and~\eqref{turing3}, that we now rewrite
\begin{eqnarray}
    a>0\quad\alpha>0, \quad
    c\gamma>\alpha D_0+aD_1, \\ 
    (c\gamma-\alpha D_0-aD_1)^2>4D_0 D_1 a\alpha,\label{eq:turing}
\end{eqnarray}
and the simplicial complex is such that $ h\in\ker \mathbf{L}_1$. 
\begin{figure*}[!htb] 
	\centering
		\includegraphics[width = 1.9 \columnwidth]{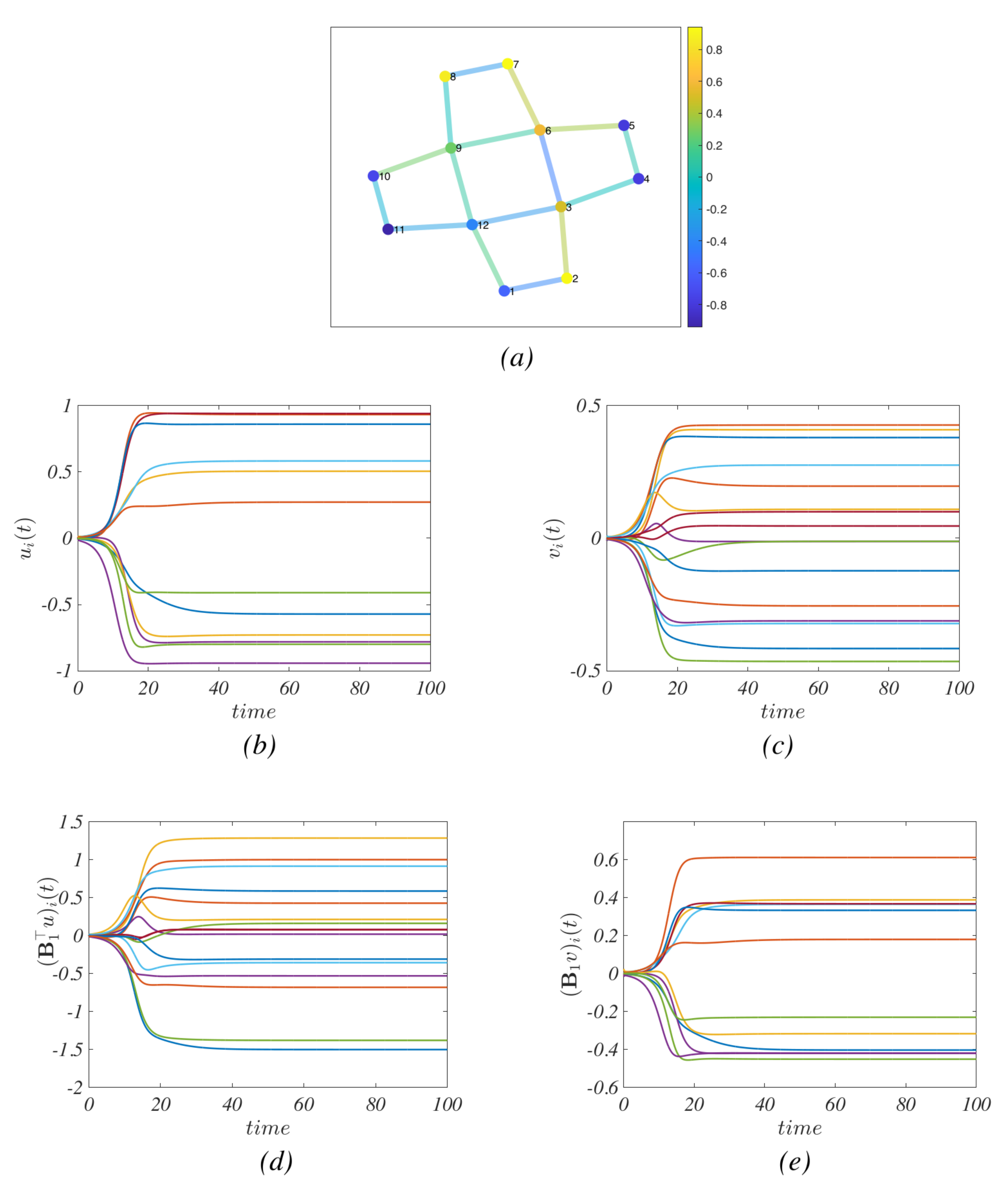}
		\caption{\textcolor{black}{$a)$ Turing patterns for species defined on nodes and on links described by model~\eqref{eq:model} on a network satisfying the conditions for the existence of an homogeneous equilibrium. In panels $b)$ and $c)$ we depict time series of the two species $u$ and $v$ on the nodes and the links, respectively, while panels $d)$ and $e)$ show the time series of the projection of the two species with the action of the boundary operator $\mathbf{B}_1$. The parameters are $a=\alpha=b=\beta=\gamma=D_0=D_1=1$ and $c=6$. The perturbation defining the initial condition, is $\sim10^{-2}$. }}\label{fig:fig1}
\end{figure*}  
\twocolumngrid
A simple example of a $1$-dimensional simplicial complex satisfying the latter condition is provided by a network of \textcolor{black}{$12$} nodes \textcolor{black}{and $16$ links}, whose nodes degrees are even and with closed loops. Note that the latter is chosen to be a subset of a square lattice.
\textcolor{black}{In Fig.~\ref{fig:fig1} we report the result of numerical simulation clearly showing the emergence of Turing patterns, namely stationary equilibria where the concentrations vary across nodes and links,  moreover the system state is far from the homogeneous solution $(u_0,v_0)=(0,0)$.}
In Fig.~\ref{fig:fig1}$.a$ the nodes and links are colored according to the asymptotic concentration of respectively $u$ and $v$ and we can thus have a geometrical view of the emerging pattern. On the other hand a dynamical view is presented in Fig.~\ref{fig:fig1}$.b-c$ where we report the nodes concentration, $u_i(t)$, and links concentration, $v_i(t)$, as a function of time \textcolor{black}{and we can observe the deviation from the homogeneous solution and the stationary asymptotic behavior of the solution}. From this figure one can clearly appreciate the onset of the instability at short time \textcolor{black}{because of the Turing condition, namely the positive dispersion relation (see Fig.~\ref{fig:fig1bis})}, pushing the initial conditions far from the equilibrium state $(u_0,v_0)=(0,0)$. 
Interestingly we observe that the projected dynamics also display a Turing pattern (see Fig.\ref{fig:fig1}.d and Fig.\ref{fig:fig1}.e).

\textcolor{black}{To have a global view, we report in Fig.~\ref{fig:fig1bis} the Turing region in the plane $(c,\gamma)$, i.e., the pairs for which the Turing instability is realized. In the main panel (B) we show the maximum of the real part of dispersion relation as a function of $c$ and $\gamma$ by using a color code, white corresponding to the impossibility of Turing instability while red to yellow are associated to the onset of the instability. The left panels, (A1), (A2) and (A3), correspond to a choice for which Turing patterns cannot emerge as confirmed by the negativity of the dispersion relation (A1) and the vanishing of the node and link amplitude (A2 and A3). The latter being defined by $A_{\mathrm{node}}(t)=\sqrt{\sum_{i=1}^{N_0}(u_i(t)-u_0)^2}$ for the nodes and $A_{\mathrm{link}}(t)=\sqrt{\sum_{j=1}^{N_1}(v_j(t)-v_0)^2}$ for the links, where $u_0$ (resp. $v_0$) is the nodes (resp. links), homogeneous equilibrium value. The right panels are associated to parameters inside the Turing region and indeed the dispersion relation assumes positive values (C1) and the node and link amplitude are strictly positive (C2 and C3). Let us observe that the amplitude can be thus considered as an order parameter capable of distinguishing between the presence or the absence of patterns.}
\begin{figure*}[!htb] 
	\centering
		\includegraphics[width = 1.9 \columnwidth]{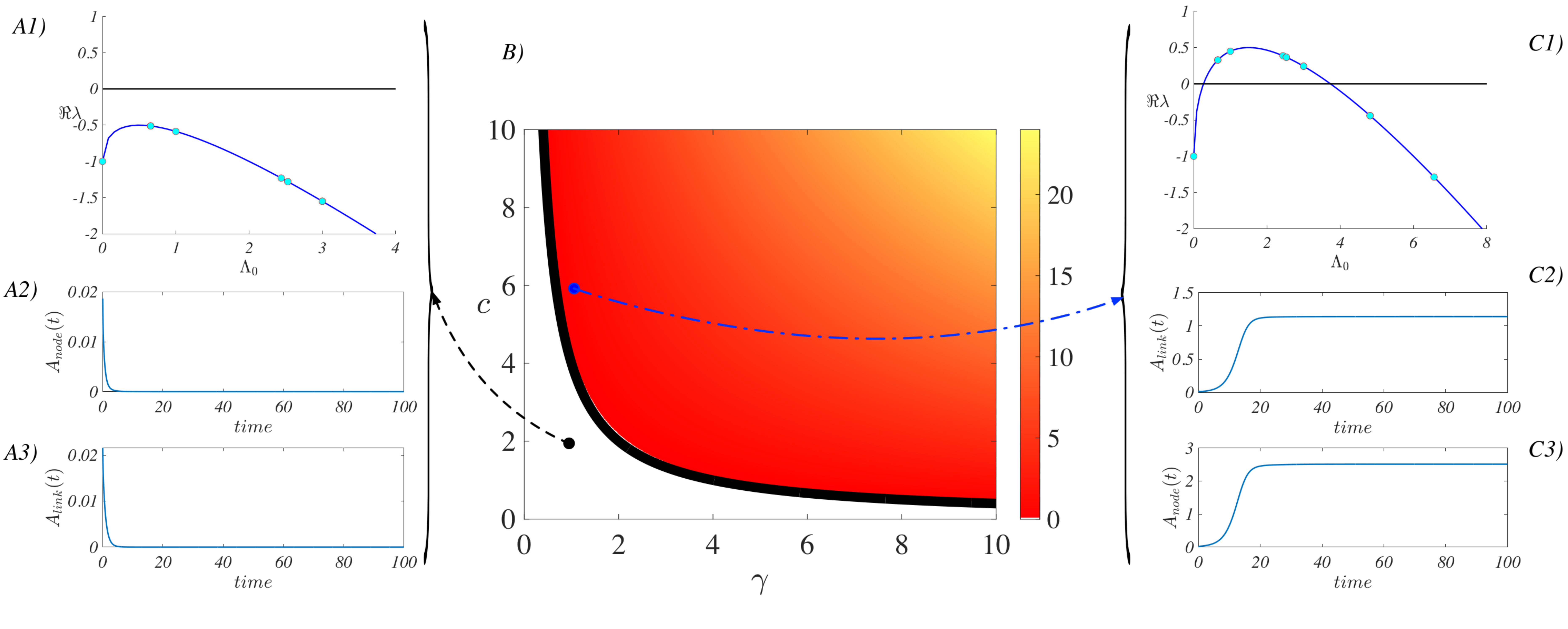}
		\caption{\textcolor{black}{Turing region in the parameters space $(c,\gamma)$. In the main panel (B), we report the region of parameters for which the Turing instability emerges; having fixed $a=\alpha=b=\beta=D_0=D_1=1$ we show the maximum of the real part of dispersion relation as a function of $c$ and $\gamma$, by using a color code (yellow corresponding to large values, red to small but positive ones and white to negative ones). The black solid curves is given by $c\gamma = \sqrt{4 D_0D_1 a\alpha}+\alpha D_0+aD_1$ (see Eq.~\eqref{eq:turing}). Panels A1), A2) and A3) correspond to the choice $(c,\gamma)=(2,2)$ that lies outside the Turing region; one can observe that the dispersion relation (panel A1) is negative and indeed patterns cannot develop as shown by the node (resp. link) amplitude (panel A2) resp. A3) decaying to $0$. Panels C1, C2 and C3) show similar results but for $(c,\gamma)=(6,2)$ inside the Turing region; the dispersion relation (C1) reaches positive values and the node (resp. link) amplitude stabilizes far from zero (C2, resp. C3).}}\label{fig:fig1bis}
\end{figure*}  

\begin{figure*}[ht] 
	\centering
		\includegraphics[width = 0.85\textwidth]{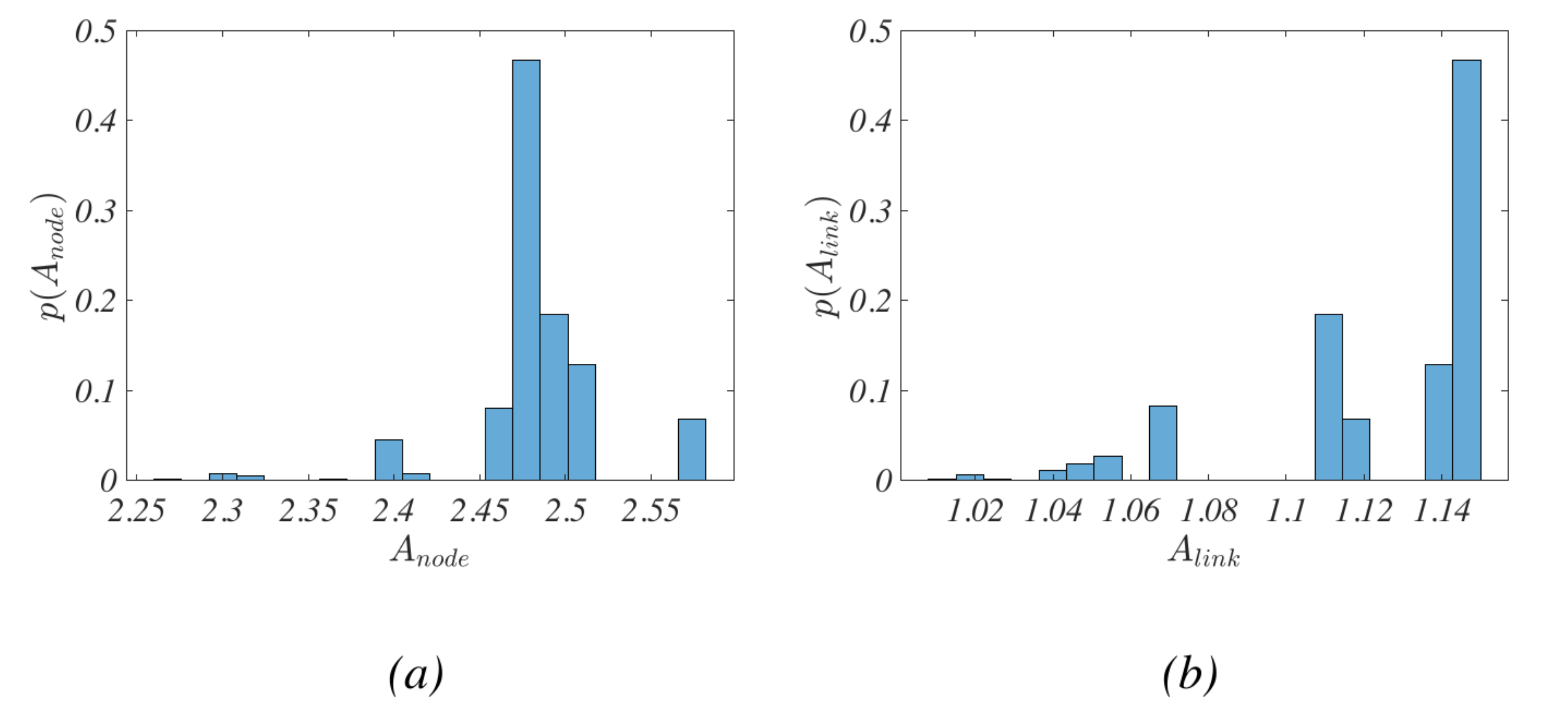}
		\caption{\textcolor{black}{We report the distribution of the node (a) and link (b) amplitude of the Turing patterns obtained by numerically simulating $5000$ times  system~\eqref{eq:model} with the parameters used in Fig.~\ref{fig:fig1} and by changing the initial conditions.}}\label{fig:fig3new}
\end{figure*}

\twocolumngrid
\textcolor{black}{Having fixed the topology of the support and the model parameters, nodes and links amplitudes depend on the initial conditions and the peculiar dynamical path followed by the system to settle into the pattern. In Fig.~\ref{fig:fig3new} we report the distribution of $A_{node}$ and $A_{link}$ once we repeat several times the numerical simulations by changing the initial conditions. We can observe that both distributions are peaked at some value and the dispersion is relatively small, however let us stress that the link amplitude distribution is very skewed.}
\twocolumngrid
\section{Interacting topological signals of nodes and links with Dirac cross-diffusion term}

We now consider the dynamics including the Dirac cross-diffusion terms. In particular we first cover the linear cross-diffusion case and leave the analysis of the cubic cross-diffusion term to a next section.

\subsection{Cross-diffusion term linear in the Dirac operator}

Topological signals on nodes and links can be coupled by a linear cross-diffusion term, leading to the reaction-diffusion dynamics
\begin{equation}\label{linear2}
\dot{\Phi}={F}(\Phi,\mathcal{D}\phi)-\tilde{\gamma}\mathcal{D}\Phi-\gamma\mathcal{L}\Phi,
\end{equation}
where the dynamical state of the network is captured by the vector $\Phi=(u,v)^{\top}$.
The diagonal  $(N_0+N_1)\times (N_0+N_1)$ matrix $\tilde{\gamma}$ of cross-diffusion constants is here chosen to have block structure 
\begin{equation}
\tilde{\gamma} = \begin{pmatrix}
D_{01}{\bf I}_{N_0} & 0 \\
0 & D_{10}{\bf I}_{N_1}
\end{pmatrix}.
\end{equation}
In particular the coupled dynamics of the topological signals $u$ and $v$ can be re-written as 
\begin{equation}\label{syst2_linera}
\begin{aligned}
			&\frac{du}{dt}={f}(u, \mathbf{B}_1 v) - D_{01} \mathbf{B}_1 v- D_0\lap_{0} u, \\
			&\frac{dv}{dt}={g}\left(v,\mathbf{B}_1^{\top} u\right)-D_{10} \mathbf{B}_1^\top u-D_1 \lap_{1}  v  \, .  
\end{aligned}
\end{equation}
In Appendix \ref{appC:linear_crossdiff} we prove that system~\eqref{linear2} can be mapped onto~\eqref{Dirac_reaction2} and thus results from the previous section can be used to derive  the conditions under which the  reaction-diffusion dynamics with the linear cross-diffusion term displays Turing patterns. Namely, the stability of the homogeneous solution~\eqref{eq:stabilityhom} and the existence of at least one unstable mode that is guaranteed by \textcolor{black}{the following two conditions to hold true}:
\begin{eqnarray}
A &=& D_0\partial_v g+D_1\partial_{u} f+(\partial_{\mathbf{B}_1^\top u} g-D_{01}) (\partial_{\mathbf{B}_1v} f-D_{10})>0\, ,\nonumber \\
A^2&>& 4D_0D_1\partial_{u} f \partial_v g\, ,
    \label{turing2b}
\end{eqnarray}

Let us stress a major consequence of these conditions, i.e., the cross-diffusion term is the driver for the instability. Indeed the cross-diffusion term enforced through the Dirac operator allows the onset of Turing patterns also in situations where patterns can never emerge if we silence cross-diffusion. In particular we can observe Turing patterns in presence of Dirac-type crossed-diffusion patterns, also when the reaction term only depends on $\Phi$ but not on $\mathcal{D}\Phi$, i.e., 
\begin{equation}
\label{eq:nodirac}
{F}(\Phi,\mathcal{D}\Phi)=F(\Phi)=\begin{pmatrix} {f}(u)\\ {g}(v)\end{pmatrix},
\end{equation} 
as long as Eq.~\eqref{eq:stabilityhom} and Eqs.~\eqref{turing2b} hold which can occur as long as $D_{01}D_{10}>0$. Let us recall that, as discussed in the previous section, under the latter assumption~\eqref{eq:nodirac}, Turing patterns cannot develop in absence of linear cross-diffusion terms. \textcolor{black}{Indeed if $D_{01}=D_{10}=0$, the variables $u_i$ and $v_i$ in system~\eqref{syst2_linera} become decoupled and thus, because of condition~\eqref{eq:stabilityhom} and the non-positivity of the spectra of $\mathbf{L}_0$ and $\mathbf{L}_1$, the homogeneous equilibrium is stable also with respect to heterogeneous perturbations}.

\textcolor{black}{Let us conclude this section by observing that Turing instability can also emerge for systems where the coupling is realized solely with the Dirac operator, namely there is no need to include the two \textcolor{black}{Hodge-Laplacian} matrices, $\mathbf{L}_0$ and $\mathbf{L}_1$ in Eq.~\eqref{syst2_linera}. This claim can be proved by simply setting $D_0=D_1=0$ into Eq.~\eqref{turing2b} and requiring thus
\begin{equation*}
(\partial_{\mathbf{B}_1^\top u} g-D_{01}) (\partial_{\mathbf{B}_1v} f-D_{10})>0\, ,
\end{equation*}
the second relation in~\eqref{turing2b} being automatically satisfied.}

\subsection{Cross-diffusion term cubic in the Dirac operator}

Cross-diffusion terms for topological signals can be also implemented with a cubic Dirac operator in the reaction-diffusion dynamics
\begin{equation}\label{cubic}
\dot{\Phi}={F}(\Phi,\mathcal{D}\phi)-\mathcal{L}(\gamma \Phi + \tilde\gamma\mathcal{D}\Phi),
\end{equation}
which can be also written in terms of the signals $u$ of the nodes and the signals $v$ of the links as
\begin{equation}\label{syst2}
\begin{aligned}
			&\frac{du}{dt}=\tilde{f}(u, \mathbf{B}_1 v) - \lap_{0} (D_0 u + D_{01} \mathbf{B}_1 v),\\
			&\frac{dv}{dt}=\tilde{g}\left(v,\mathbf{B}_1^{\top} u\right)- \lap_{1} (D_1 v + D_{10} \mathbf{B}_1^\top u) \, .
\end{aligned}
\end{equation}
Starting  from the existence of a homogeneous equilibrium $(u^*$, $v^*)$ that we assume to be stable with respect to homogeneous perturbations, we can  determine the conditions for the onset of Turing instability. We thus consider perturbations about such equilibrium, $\delta u=u-u^*$, $\delta v=v-v^*$, whose evolution is given by the linearized system
\begin{equation}	
\begin{aligned}
	&\frac{d\delta u}{dt}=\left(\partial_{u} {f}\right) \delta u+\left(\partial_{\mathbf{B}_1v} {f}\right) \mathbf{B}_1 \delta v-\lap_{0}(D_0\delta u +D_{01}\mathbf{B}_1 \delta v ),\\
	&\frac{d\delta v}{dt}=(\partial_{\mathbf{B}_1^\top u}{g}) \mathbf{B}_1^\top \delta u+(\partial_v {g}) \delta v-\lap_{1}(D_1\delta v +D_{10}\mathbf{B}_1^\top\delta u)\, . 
\end{aligned}
\end{equation}
Considering the stability of perturbations within the kernel of the Laplacians leads to the stability  conditions given by Eq.~\eqref{eq:stabilityhom}, \textcolor{black}{because of the assumption $\mathbf{B}_1 h = \mathbf{B}_1^\top h=0$ where $h=(1,...,1)^\top$.}

On the other hand, \textcolor{black}{by} considering a generic perturbation and projecting it on the \textcolor{black}{Laplacian} eigenbasis, we obtain a new Jacobian matrix, $\mathcal{J}_k$
\begin{equation}
\mathcal{J}_k=	\left(\begin{matrix} 
	\partial_{u} {f}- D_0b_k^2  &b_k \partial_{\mathbf{B}_1v} {f} - D_{01} b_k^3\\
	b_k\partial_{\mathbf{B}_1^\top u} {g} - D_{10} b_k^3 &  \partial_{v} g- D_1 b_k^2 
\end{matrix}\right)\, ,
\end{equation}
whose spectrum determines the stability of the heterogeneous perturbation and thus the possible onset of the instability.

The eigenvalues of $\mathcal{J}_k$ are determined by solving
\begin{equation}
	\det\left(\begin{matrix} 
	\partial_{u} f- D_0b_k^2 -\lambda &b_k \partial_{\mathbf{B}_1v} f - D_{01} b_k^3\\
	b_k\partial_{\mathbf{B}_1^\top u} g - D_{10} b_k^3 &  \partial_{v} g- D_1 b_k^2 -\lambda
\end{matrix}\right)=0\, ,
\end{equation}
which can  be rewritten as
\begin{equation}
\label{eq:lambdaXdiff}
	\lambda^{2}+\lambda \tilde{\Gamma}_{1}\left(b_k^{2}\right)+\tilde{\Gamma}_{2}\left(b_k^{2}\right)=0\, ,
\end{equation}
where $\tilde{\Gamma}_1(b_k^2)=\Gamma_1(b_k^2)$ is given by Eq.(\ref{eq:term0}) and is then always positive if the homogeneous equilibrium is stable. In this scenario $\tilde{\Gamma}_2(b_k^2)$ is a cubic polynomial in $b_k^2$, given by
\begin{equation}
\tilde{\Gamma}_2(b_k^2) =  \tilde{a}_3 b_k^6+\tilde{a}_2 b_k^4+\tilde{a}_1 b_k^2+ \tilde{a}_0,
\end{equation}
with 
\begin{eqnarray}
\tilde{a}_3&=&-D_{01} D_{10},\\
\tilde{a}_2&=&(D_0 D_1 +D_{01} \partial_{\mathbf{B}_1^\top u} g + D_{10} \partial_{\mathbf{B}_1 v} f ),
\end{eqnarray}
and $\tilde{a}_1=a_1,\tilde{a}_0=a_0$.
As for the case without cross-diffusion, also in this setting, only stationary Turing patterns can be observed.

We consider exclusively the situation in which we have  $D_{01}D_{10}<0$  which enforces the stability of modes corresponding to large values of $\Lambda_0$.
In this case, the conditions to observe stationary Turing patterns are, in addition to~\eqref{eq:stabilityhom}, that one of the two following inequalities needs to be satisfied
\begin{eqnarray}\label{cond_cross1m}
\tilde{a}_2&=&D_0 D_1 +D_{01} \partial_{\mathbf{B}_1^\top u} g + D_{10} \partial_{\mathbf{B}_1 v} f  < 0,\nonumber \\ \tilde{a}_1&=&-\left(D_1\partial_{u} f+ D_0\partial_v g+\partial_{\mathbf{B}_1^\top u} g\ \partial_{\mathbf{B}_1v} f\right)<0, \label{cond_cross1bm}
\end{eqnarray}
together with 
{
\begin{equation}
 2D_0D_1 K_++D_0^2D_1^2 +K_-^2>0,
\label{cond_cross2bm}   
\end{equation}
where $K_{\pm}$ is given by 
\begin{equation}
K_{\pm}=D_{01}\partial_{{\bf B}_1^\top u}g\pm D_{10}\partial_{{\bf B}_1v}f
\end{equation}
}
(see Appendix \ref{appD:conditions_cubic} for the derivation of these results).

Interestingly, from  this study it emerges that for a cubic Dirac cross-diffusion term, as long as $D_{01}D_{10}<0$ we cannot observe the onset of the Turing instability for a reaction term of the type $F(\Phi,\mathcal{D}\Phi)=F(\Phi)$. Indeed in this case we have $\partial_{{\bf B}_1^\top u} g=0$ and $\partial_{{\bf B}_1 v} f=0$ and hence neither one of the conditions $\eqref{cond_cross1bm}$ can be satisfied when  the stability condition \eqref{eq:stabilityhom} holds.
 \begin{figure*}[h!] 
	\centering
		\includegraphics[width = 0.85\textwidth]{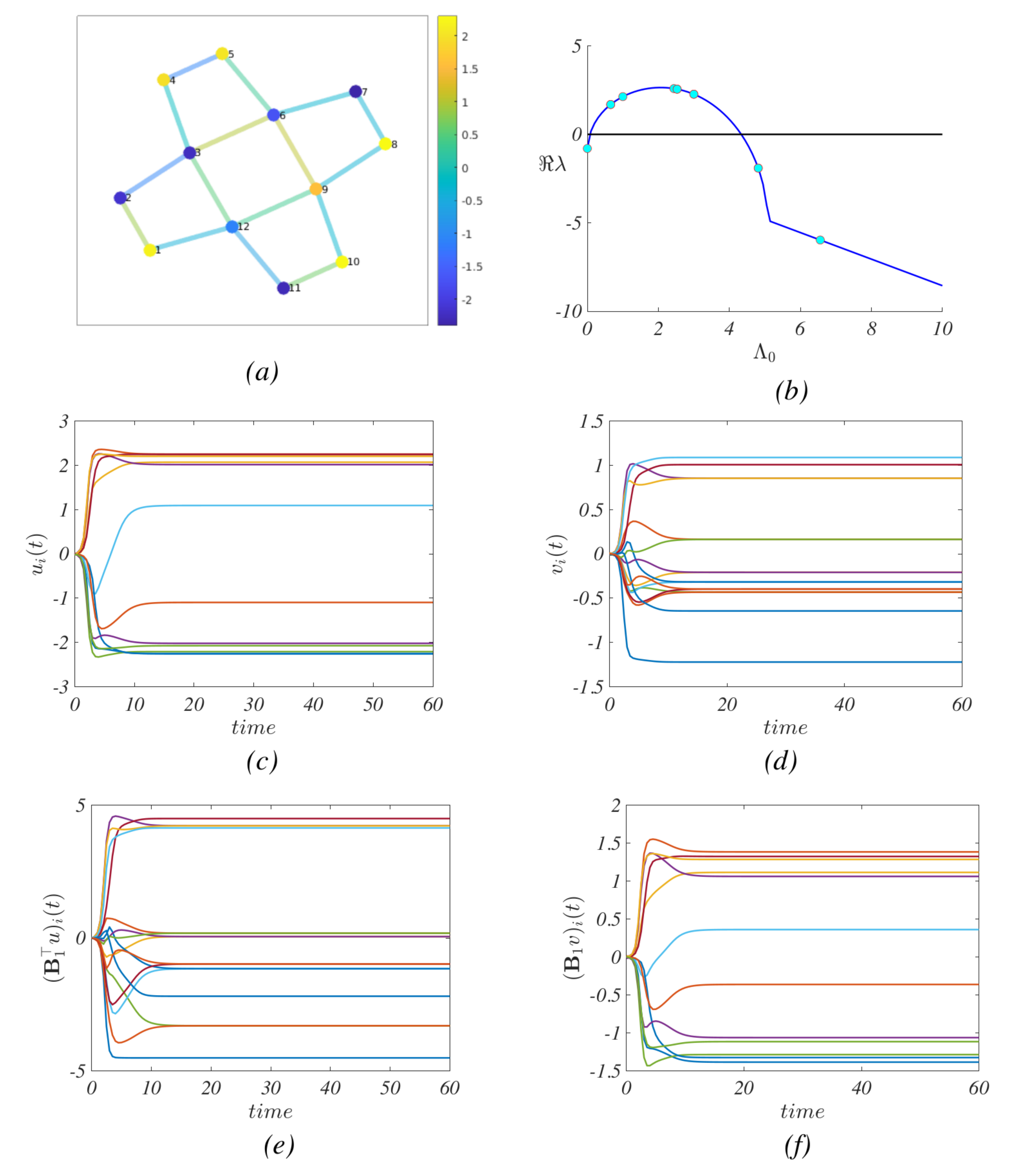}
		\caption{$a)$ Turing patterns for the species on the nodes and on the links described by model \eqref{eq:model2} on a network satisfying the conditions for the homogeneous equilibrium; $b)$ dispersion relation: in blue we depict the continuous curve, computed \textcolor{black}{by replacing the discrete parameter $b^2_k$ with a continuous variable}, while the cyan dots are the actual dispersion relation, where now the onset of the Turing instability is a function of the (real) spectrum of $\lap_0$\textcolor{black}{, i.e., computed by using the discrete values of $b_k^2$}. In panels $c)$ and $d)$ we depict time series of the two species $u$ and $v$ on the nodes and the links, respectively, while panels $e)$ and $f)$ show the time series of the projection of the two species with the action of the boundary operator $B_1$. The parameters are $a=0.8$, $\alpha=1.3$, $b=1$, $\beta=0.5$, $c=8$, $\gamma=2$, $D_0=0.5$, $D_1=1$, $D_{01}=-1.5$ and $D_{10}=0.4$; the initial perturbation is $\sim10^{-2}$.}\label{fig:fig2}
\end{figure*}

\subsection{Numerical results with a cubic Dirac cross-diffusion term}
Let us now numerically validate the above analysis of the reaction-diffusion system with cubic Dirac cross-diffusion terms. By considering  the benchmark model \eqref{eq:model} with the addition of cubic Dirac cross-diffusion terms, we obtain
 \begin{eqnarray}
  \dot{u}&=&-a u-bu^3+c \mathbf{B}_1 v - \lap_0(D_0 u+D_{01}\mathbf{B}_1 v), \nonumber \\ 
 \dot{v}&=& -\alpha v-\beta v^3 +\gamma \mathbf{B}_1^{\top}u - \lap_1(D_1 v+ D_{10}\mathbf{B}_1^\top u). \label{eq:model2}
 \end{eqnarray}
Let us assume conditions~\eqref{eq:stabilityhom},~\eqref{cond_cross1bm} and~\eqref{cond_cross2bm} to hold true, and, to be concrete, let us consider the case $D_{01}<0$ and $D_{10}>0$. Assuming once again \textcolor{black}{to} work with the simplicial complex used in the previous section, \textcolor{black}{then} Turing patterns can emerge as shown in Fig.~\ref{fig:fig2}$.a$, where nodes and links, colored according to the asymptotic values of $u_i$ and $v_i$, \textcolor{black}{clearly show a dependence of the solution on the latter ones}. In Fig.~\ref{fig:fig2}$.c-e$ we report the temporal evolution of $u_i(t)$, $v_i(t)$ \textcolor{black}{and one can clearly appreciate how far from the homogeneous state they are; a similar result can be observed for} their projections ${\bf B}_1^{\top}u$ and ${\bf B}_1v$. \textcolor{black}{Finally,} the dispersion relation is presented in Fig.~\ref{fig:fig2}$.b$ \textcolor{black}{to support the claim of short time instability}.

\section{Conclusions}

In this paper we have formulated reaction-diffusion dynamics of topological signals defined on nodes, links, and higher-order simplices of simplicial complexes \textcolor{black}{or cells of cell complexes}. In this framework, each  species of reactants lives on simplicies \textcolor{black}{or cells} of a given dimension, for instance in a simplicial complex of dimension $d=2$ one would consider three \textcolor{black}{kind of} species living on nodes, links and triangles.
Species associated to simplices of different dimension can be coupled thanks to the Dirac operator which projects a signal defined on $n$-dimensional simplices either one dimension up or one dimension down.
In the proposed reaction-diffusion dynamics, the coupling can then be enforced either by a Dirac reaction term or/and Dirac cross-diffusion terms.
After discussing the general framework valid for simplicial \textcolor{black}{and cell} complexes of arbitrary dimension, we focus on the reaction-diffusion dynamics of topological signals defined on networks, i.e., coupling the dynamics between links and nodes, and we establish conditions for the onset of the Turing instability. The latter conditions are derived when signals of different dimension are only coupled with the Dirac reaction term, as well as when they are also coupled by a linear or a cubic Dirac cross-diffusion term.

We have found that the Turing patterns arising from the reaction-diffusion dynamics of topological signals are never localized only on nodes or links of the network. Instead they always involve both node and link signals. Moreover, the projection of the link signals on the nodes, and the projection of the node signals onto the links are shown to also display a Turing pattern.

We also observe that when the reaction term does not depend on the projected signal, the Turing pattern can be observed only in presence of a linear Dirac cross-diffusion term. 

Our results are validated on a small toy model for the reaction-diffusion of topological signal on a network, and on simulations of square lattices with periodic boundary conditions.

\section*{Acknowledgements}
R.M. is supported by a FRIA-FNRS PhD fellowship, Grant FC 33443, funded by the Walloon region. G.B. acknowledges support from the Royal Society (IEC\textbackslash NSFC\textbackslash191147).

\bibliographystyle{apsrev4-1}
\bibliography{biblio.bib}

\appendix

\section{Basics properties of algebraic topology}

\label{appA:alg_topology}
\setcounter{equation}{0}
\renewcommand{\theequation}{A\arabic{equation}}
\setcounter{figure}{0}
\renewcommand{\thefigure}{A\arabic{figure}}
\subsubsection*{Simplicial \textcolor{black}{and cell} complexes, the boundary and coboundary operators}
    
	\textcolor{black}{A $d$-dimensional cell complex $\mathcal{S}$ is a collection of cells whose dimension $n$ is smaller or equal to $d$ which is closed under the inclusion of the cells' faces. The $n$-dimensional cells are convex polytopes of dimension $n$, i.e., for $n=0$ they are nodes, for $n=1$ they are links, for $n=2$ they are triangles, squares, pentagons etc. and for $n=3$ they are tetrahedra, hypercubes, orthoplex etc.
	The faces of an  $n$-cell are the $(n-1)$-dimensional cells at its boundary.  A special case of cell complex is a simplicial complex which is only formed by simplices, i.e., cells whose underlying network structure is a clique, such as  nodes, links, triangles, tetrahedra and so on.
	The cells of a cell complex are {\em oriented} and typically for simplicial complexes the orientation of the simplicial complex induced by the nodes label is used, for instance a link $[i,j]$ is positively oriented if $i<j$ and similarly a triangle $[i,j,k]$ and all the triangles obtained by a  cyclic permutation of the indices are positively oriented if $i<j<k$. For more information about simplicial and cell complexes see Refs.\cite{bianconi2021higher,battiston2022higher,hatcher2002algebraic}.}

\begin{figure}[H]
	\centering
		\includegraphics[width = 0.4\textwidth, angle=0]{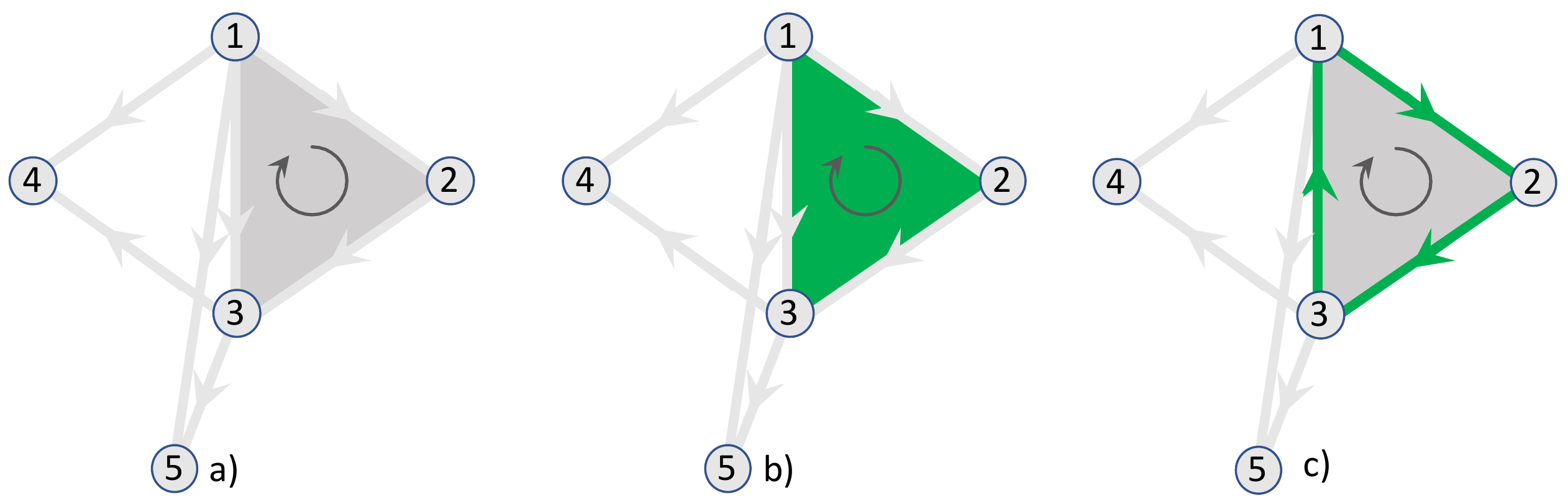}
	\caption{Panel a) shows a simplicial complex on dimension 2, with simplicial orientation induced by a labeling of the nodes. The boundary of the 2-simplex $[1,2,3]$ highlighted in panel Panel b) is shown in panel c)}
	\label{f:fig1v2}
\end{figure}
The topology of cell complexes can be investigated using methods coming from algebraic topology.
Let us indicate with $N_m$ the number of $m$-dimensional cells present in the considered cell complex.
In algebraic topology the cells $\mu_n^{(m)}$ of dimension $n$ of a simplicial complex define the basis of a vector space $C_n$ of  $n$-\textit{chains}. Therefore a $n$-chain ${\bf c}\in C_n$ is a finite linear combination of the  $n$-cells $\mu^{(m)}_n$ with $1\leq m\leq N_{n}$ with coefficients $c_i$
	\begin{equation}\label{key}
		{\bf c} = \sum_{m=1}^{N_n} c_i\mu^{(m)}_n\, .
	\end{equation} 
    The boundary of a chain can be obtained from a chain by applying to it the \textit{boundary operator}  $\partial_n :C_n\rightarrow C_{n-1}$ which is represented by the boundary matrix ${\bf B}_{n}$

The boundary matrix ${\bf B}_n$ is a $N_{n-1}\times N_{n}$ rectangular matrix of elements $[B_{n}]_{\mu,\mu'} = +1$ if $\mu$ is a $(n-1)$-dimensional face of the \textcolor{black}{$n$-cell} $\mu$ with coherent orientation, $[B_{n}]_{\mu,\mu'} = -1$ if the orientation is not coherent, and $[B_{n}]_{\mu,\mu'} = 0$ if $\mu$ is not a face of $\mu'$. In the particular case of $\mathbf{B}_{1}$, we have for instance
    \begin{equation}
    [B_{1}]_{i\ell}=\left\{\begin{array}{ccc}1 & \mbox{if}\   \ell=[j,i]\  \mbox{and}\   j<i,\\ -1 & \mbox{if} \  \ell=[i,j] \  \mbox{and}\   i<j,\\  0 & \  \mbox{otherwise}\end{array}\right.
    \end{equation}
    once we assume the orientation to be induced by the nodes labels. 
    As an example, the matrix $\mathbf{B}_1$ and $\mathbf{B}_2$ of the simplicial complex shown in Fig.~\ref{f:fig1v2} are given by
\begin{align*}
  {\bf B}_1 = \begin{blockarray}{l c c c c c c c}
                       & [1,2] & [1,3] & [1,4] & [1,5] & [2,3] & [3,4] & [3,5]\\[-0.6ex]
  \begin{block}{l [c c c c c c c]}
   \text{[1]} & -1 & -1 & -1 & -1& 0 & 0& 0\topstrut \\
   \text{[2]} & 1 & 0 & 0 &0& -1 & 0 &0\\
   \text{[3]} & 0 & 1 & 0 &0& 1 & -1&-1 \\
   \text{[4]} & 0 & 0 & 1 &0& 0 & 1&0\\
   \text{[5]} & 0 & 0 & 0 &1& 0 & 0&1\botstrut \\
  \end{block}
  \end{blockarray}
\end{align*}
and
\begin{align*}
  {\bf B}_2 = \begin{blockarray}{l c}
                       & [1,2,3] \\[-0.6ex]
  \begin{block}{l [c]}
  \text{[1,2]} & 1 \topstrut \\
  \text{[1,3]} & -1 \\
  \text{[1,4]} & 0 \\
  \text{[1,5]} & 0\\
  \text{[2,3]} & 1 \\
  \text{[3,4]} & 0 \\
  \text{[3,5]} &0 \botstrut \\
  \end{block}
  \end{blockarray}
\end{align*}
 The set of all the boundary matrices ${\bf B}_n$ with $0\leq n\leq d$ of a simplicial complex fully encodes the topology of the simplicial complex.
 The adjoint of the boundary operator or {\em coboundary operator} is represented by the matrix ${\bf B}_n^\top$. 

 \subsubsection*{The Hodge-Laplacians and the Dirac operator of a cell complex}
 Starting from the boundary and the coboundary operators, we define the higher-order Laplacians and the Dirac operator.
 
The Laplace operator \cite{josthorak2013spectra,lim2020hodge,bianconi2021higher} of order $n$, also called $n$-Hodge-Laplacian, describes higher-order diffusion from \textcolor{black}{$n$-cells to $n$-cells} and are $N_n \times N_n$ matrices defined as 
	\begin{equation}\label{hodgeLap}
		\mathbf{L}_{n}=\mathbf{B}_{n}^{\top} \mathbf{B}_{n}+\mathbf{B}_{n+1} \mathbf{B}_{n+1}^{\top}= \lap^{down}_n+\lap^{up}_n
	\end{equation}
    for $1\leq n<d$. For $n=0$ and $n=d$ the Hodge-Laplacians $\mathbf{L}_{0}$ and $\mathbf{L}_{d}$ are respectively given by $\mathbf{L}_{0} = \lap^{up}_0 = \mathbf{B}_{1} \mathbf{B}_{1}^{\top}$ and $\mathbf{L}_{d} = \lap^{down}_d = \mathbf{B}_{d}^\top \mathbf{B}_{d}$.
    
    The action of the Hodge-Laplacian can be interpreted as follows. The term $ \lap^{up}_n $ represents the diffusion between \textcolor{black}{$n$-cells} through shared \textcolor{black}{$(n+1) $-dimensional cells}. In the case of a network, as previously noticed, this is the combinatorial Laplacian, where concentrations on nodes diffuse through incident links. The term $\lap^{down}_n$  represents diffusion between \textcolor{black}{$n$-cells through shared $(n-1)$-cells}, i.e., incident $(n-1)$-faces.
For instance in a network (i.e., a $1$-simplicial complex) ${\bf L}_1={\bf L}_1^{down}$ determines diffusion from links to links through nodes.

From this definition, it is clear that the Hodge-Laplacian of order $n$ only acts on topological signals of dimension $n$. Therefore the $n$-Hodge-Laplacian cannot couple signals of different dimension.

In order to couple signal of different dimension, we require the Dirac operator \cite{bianconi2021topological,lloyd2016quantum,ameneyro2022quantum,post2009first}, $\mathcal{D}$, which is encoded by an $M\times M$ matrix where $M=\sum_{n=0}^d N_n$ and has elements
\begin{eqnarray}
    \mathcal{D}_{\mu,\mu'}=\left\{\begin{array}{ccc}[B_n]_{\mu,\mu'} & \mbox{if } & |\mu'|=|\mu|+1=n\\
    {[B_n^{\top}]}_{\mu,\mu'} 
    & \mbox{if } & |\mu|=|\mu'|+1=n
    \end{array}\right.,
\end{eqnarray}
where with $|\mu|$ we indicate the dimension of the \textcolor{black}{cell} $\mu$.
It follows that in a two dimensional \textcolor{black}{cell} complex, the Dirac operator has the block structure 
\begin{equation}
\mathcal{D}=\begin{pmatrix}0&{\bf B}_1&0\\
{\bf B}_1^{\top}&0&{\bf B}_2\\
0&{\bf B}_2^{\top}&0\end{pmatrix},
\end{equation}
while in a network the Dirac operator is given by 
\begin{equation}
\mathcal{D}=\begin{pmatrix}0&{\bf B}_1\\
{\bf B}_1^{\top}&0\end{pmatrix},
\end{equation}
It follows that the Dirac operator, differently from the Hodge-Laplacian, can couple topological signals of different dimension.
In particular the Dirac operator can be used to project a topological signal of any dimension $n$ onto simplices of dimension $n+1$ and $n-1$.
One of the most significant properties of the Dirac operator is that it can be considered the ``square root" of the Laplacian. In fact we have 
\begin{equation}
\mathcal{D}^2=\mathcal{L}=\lap_0\oplus\lap_1\oplus\ldots\oplus \lap_d.
    \end{equation}
    For instance, for a simplicial complex of dimension $d=2$ we have 
    \begin{equation}
    \mathcal{D}^2=\mathcal{L}=\begin{pmatrix}\lap_0&0&0\\0&\lap_1&0\\0&0&\lap_2\end{pmatrix},
    \end{equation}
    and for a network 
    \begin{equation}
    \mathcal{D}^2=\mathcal{L}=\begin{pmatrix}\lap_0&0\\0&\lap_1\end{pmatrix}.
    \end{equation}
	Interestingly, both the Hodge-Laplacians and the Dirac operator can be extended to treat weighted simplicial complexes (see for instance \cite{baccini2022weighted}).
\subsubsection*{Major Spectral properties of the boundary operators, the Hodge-Laplacians and the Dirac operator} 

The $n$-order Hodge-Laplacian \cite{josthorak2013spectra,lim2020hodge,bianconi2021higher} is a semi-definite positive operator whose kernel has dimension equal to  the $n$-th Betti number $\beta_n$, i.e., the degeneracy of its null eigenvalue is equal to the Betti number $\beta_n$.
In addition to this, the Hodge-Laplacians obey the Hodge decomposition which implies that the space of $n$-chains can be decomposed as 
\begin{equation}
    C_n=\mbox{im}({\bf B}_{n}^\top)\oplus\mbox{ker}(\lap_n)\oplus\mbox{im}({\bf B}_{n+1}),
\end{equation}
where the kernel of the Hodge-Laplacians are given by 
	\begin{equation}\label{kernel}
		\operatorname{ker}(\lap_0) = \operatorname{ker}(\bnd_{1}^\top) \quad \operatorname{ker}(\lap_n) = \operatorname{ker}(\bnd_{n}) \cap \operatorname{ker}(\bnd_{{n+1}}^\top)\, .
	\end{equation}

The Dirac operator \cite{bianconi2021topological} has a kernel given by the direct sum of the kernels of the Laplacians, 
\begin{equation}
    \mbox{ker}(\mathcal{D})=\mbox{ker}(\mathcal{L})=\mbox{ker}(\lap_0)\oplus\mbox{ker}(\lap_1)\oplus\ldots\oplus \mbox{ker}(\lap_d).
\end{equation}
The non-zero spectrum of the Dirac operator is formed by the concatenation of the spectra of the Hodge-Laplacians taken with positive and negative sign.

Let us now focus on the spectrum of the Hodge-Laplacians $\lap_0$ and $\lap_1$ defined on a network, and reveal the relation between their spectrum and the singular values of the boundary operator ${\bf B}_1$. 
Since $\lap_0={\bf B}_1{\bf B}_1^{\top}$ and $\lap_1={\bf B}_1^{\top}{\bf B}_1$ it follows that $\lap_0$ and $\lap_1$ are isospectral, i.e., they have the same non-zero eigenvalues and any eigenvalue $\Lambda_0^k$ of $\lap_0$ can be written as $\Lambda_0^k=b_k^2$ where $b_k$ indicates the non-zero singular eigenvalues of the boundary matrix ${\bf B}_1$.
We note however that the degeneracy of the zero eigenvalue \textcolor{black}{$\Lambda^1_a=0$, $a=0,1$} is different for $\lap_0$ and $\lap_1$. Indeed, for $\lap_0$ the degeneracy of the zero eigenvalue is $\beta_0$, i.e., the number of connected components of the network, while for $\lap_1$ it is given by $\beta_1$, i.e., the number of independent cycles of the network. Therefore for a network that has the topology of a linear chain with periodic boundary conditions when $N_0=N_1$, we have $\beta_0=\beta_1=1$, for a tree when we have $N_1=N_0-1$ we have $\beta_0=1$ and $\beta_1=0$ and in general for a connected network we have $\beta_0=1$, $\beta_1=N_1-N_0+1$.
Let us denote by $ {\psi^{k}_a}$ the eigenvector of $\lap_{a}$ associated to the non-zero eigenvalue $\Lambda_{a}^{k}$, for $a=0,1$, namely $\lap_{a} {\psi^{k}_a} = \Lambda_a^{k}  {\psi^{k}_a}$, then we have by the properties of the singular value decomposition applied to $\mathbf{B}_1$ that 
\begin{equation}
\mathbf{B}_1\psi_1^k=b_k\psi_0^k,\quad
\mathbf{B}_1^{\top}\psi_0^k=b_k\psi_1^k\, .
\end{equation}

\section{Square Lattice with periodic boundary conditions}\label{appB:square_lattice}
\setcounter{equation}{0}
\renewcommand{\theequation}{B\arabic{equation}}
\setcounter{figure}{0}
\renewcommand{\thefigure}{B\arabic{figure}}
For this case of interest, where the \textcolor{black}{cell} complex is a $d$-dimensional square lattice with periodic boundary conditions (p.b.c.), interesting phenomena occur. For a rectangular portion of a $d$-dimensional square lattice with linear size $L_m$ in the direction $m$, the  eigenvalues and the eigenvectors of the graph Laplacians $\lap_0$ and $\lap_1$ can be easily computed \cite{bianconi2021higher}.
Indeed the eigenvectors of $\lap_0$ are the Fourier modes of the lattice associated with wave number ${\bf q}=(q_1,q_2,\ldots, q_m,\ldots, q_d)$ and the eigenvalues of $\lap_0$ can be expressed as
\begin{equation}
\Lambda_0\textcolor{black}{({\mathbf q})}=4\sum_{m=1}^d\sin^2 (q_m/2).
\end{equation}
The periodic boundary conditions impose 
\begin{equation}
q_m=\frac{2\pi}{L_m}\hat{n}_m\quad \hat{n}_m=0,1,2\ldots L_m-1.
\end{equation}

The analysis that has been carried out for the case presented in the main text will let us \textcolor{black}{to} conclude that as soon as an eigenvalue returns a positive dispersion law, i.e., $\lambda(b_k^2) > 0$, the corresponding eigenspace will be constituted by one periodic eigenvector that spans the nodes and one, again periodic, that spans the links. Consequently, as in the general case, the arising instability cannot be confined to the space of nodes or links, here too.

\begin{figure*}[h]
\centering
   	\includegraphics[width = 0.85\textwidth]{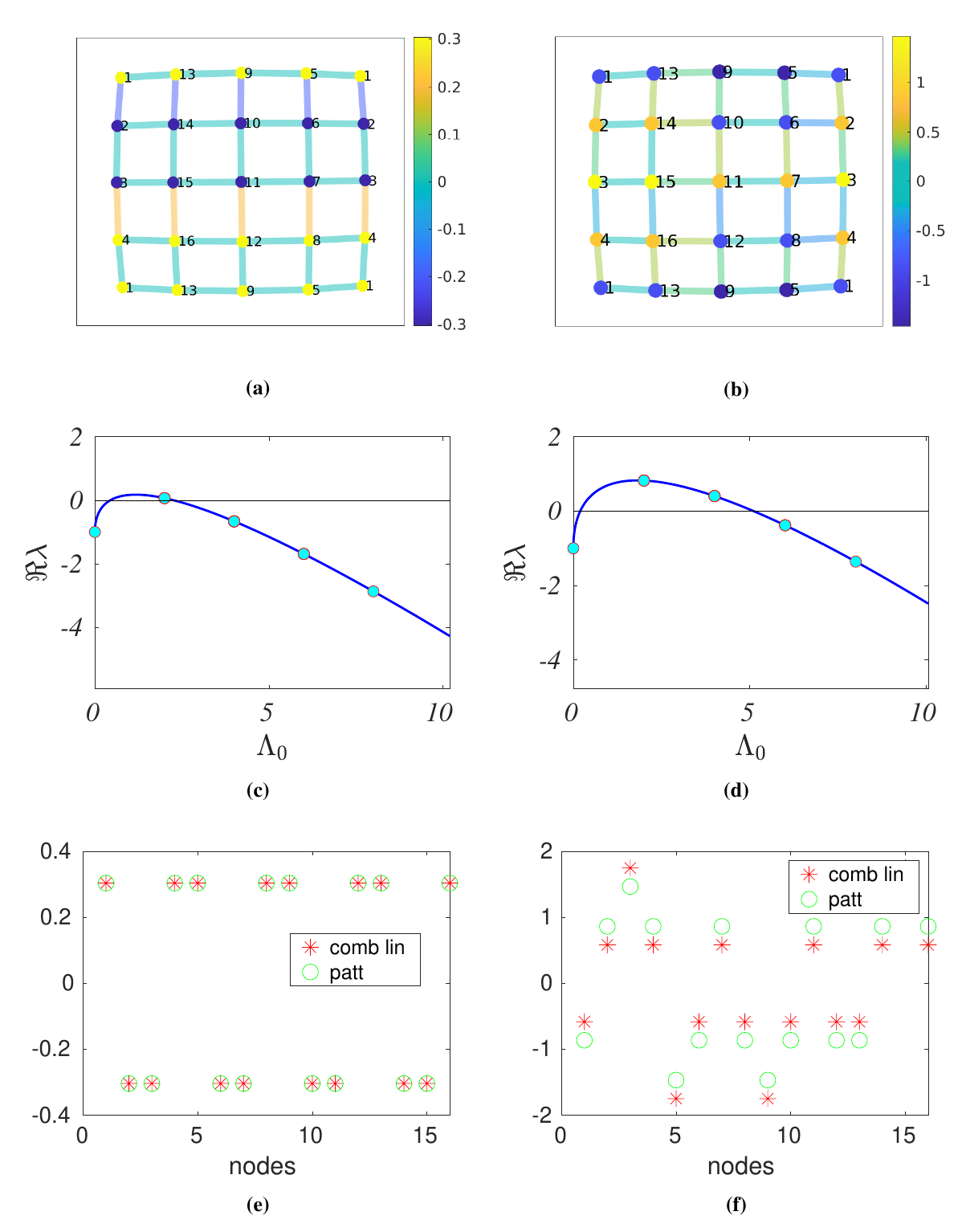}
\caption{Model \eqref{eq:model} on a $4\times 4$ $2$-dimensional lattice with p.b.c. The periodicity of lattice, shown in panels $a)$ and $b)$, is represented by adding one column and one row, so that the displayed nodes are $25$, but effectively they are $16$. Panels on the left show the case where only one mode contributes to the instability, while those on the right where multiple modes are unstable, as shown by the dispersion laws in panels $c)$ and $d)$, respectively. When only one mode is unstable, the nodes' pattern is striped-like, while the signal on the links is non-zero only when the given link connects two nodes with different signals, as shown in panel $a)$; on the other hand, when multiple modes are unstable, such regular structure is lost, as we can see in panel $b)$. Panels $e)$ and $f)$ show a comparison of the nodes' pattern with a linear combination of the eigenvector associated to the critical mode(s), showing a good accordance. The model parameters are $a=\alpha=b=\beta=\gamma=D_0=D_1=1$; $c=4.7$ for panels $a),~c),~e)$, while $c=7.3$ for panels $b),~d),~f)$; the initial perturbation is $\sim10^{-2}$.}
\label{fig:D}
\end{figure*}

To be concrete, we have numerically analyzed the dynamical system~\eqref{eq:model} defined on a $4\times 4$ $2$-dimensional lattice with p.b.c.. The results are depicted in Fig.~\ref{fig:D}: on the left column, panels $a)$, $c)$ and $e)$, refer to the case where there is single unstable mode, while on the right column (panels $b)$, $d)$ and $f)$), multiple unstable modes are allowed. Let us first observe that the critical mode, i.e., the one associated to the largest value of the dispersion relation, is the same for both parameter configurations; we also remark that to each mode, except for the $0$-th one, there are associated several linearly independent eigenvectors, four vectors in the case of a $4\times 4$ lattice with p.b.c..
When only one mode is unstable (see panel c) in Fig.~\ref{fig:D}), we observe that the signal on the nodes exhibits a (horizontal) striped-like pattern and the signals on the links are non-zero only when the link connects nodes with different signals values (Fig.~\ref{fig:D}$a$). Such ordered structure is destroyed when multiple modes are unstable (Fig.~\ref{fig:D}$b,d$). When there is a single unstable mode, the stationary pattern is a linear combination of the $4$ eigenvectors associated to such mode (Fig.~\ref{fig:D}$e$); remarkably this continues to be true when more than one mode is unstable (Fig.~\ref{fig:D}$f$).

Let us conclude by observing that the former result is a slight generalization of the one we can found in~\cite{NM2010}, where authors showed that in the case of a unique unstable and non-degenerate mode, the patterns can be described by such eigenvector, despite the fact that they are the reflex of a nonlinear process. Here we have shown that the same result holds true if the unique critical eigenvalue possesses a high-dimensional subspace spanned by several eigenvectors and even in the case of multiple unstable modes.

\section{Derivation of the conditions for the onset of Turing pattern in presence of a linear Dirac cross-diffusion term}\label{appC:linear_crossdiff}
\setcounter{equation}{0}
\renewcommand{\theequation}{C\arabic{equation}}
\setcounter{figure}{0}
\renewcommand{\thefigure}{C\arabic{figure}}
In this Appendix our goal is to derive the condition for the onset of the Turing instability for the reaction-diffusion dynamics with linear Dirac cross-diffusion term which we rewrite here for convenience,
\begin{equation}\label{linear2c}
\dot{\Phi}={F}(\Phi,\mathcal{D}\Phi)-\tilde{\gamma}\mathcal{D}\Phi-\gamma\mathcal{L}\Phi.
    \end{equation}

We notice that by putting
\begin{equation}
F(\Phi,\mathcal{D}\Phi)-\tilde{\gamma}\mathcal{D}\Phi= \tilde{F}(\Phi,\mathcal{D}\Phi),
    \end{equation}
Eq.(\ref{linear2c}) reduces to Eq.(\ref{Dirac_reaction2}) with Dirac reaction term given by $\tilde{F}(\Phi,\mathcal{D}\Phi)$, i.e. it reduces to 
\begin{equation}
\dot{\Phi}=\tilde{F}(\Phi,\mathcal{D}\Phi)-\gamma\mathcal{L}\Phi.
    \end{equation}
It follows that the conditions for the onset of the Turing instability can be obtained directly from     
  Eq. \eqref{turing2} and Eq. \eqref{turing3} by making the substitutions
\begin{equation}
        \partial_{\bnd_1v}f \rightarrow \partial_{\bnd_1v}f-D_{01}, \quad \partial_{\bnd^\top_1u}g \rightarrow \partial_{\bnd^\top_1u}g-D_{10}.
\end{equation}
This allows us to obtain that in the case with linear Dirac cross-diffusion terms we can observe the onset of the Turing instability when in addition to Eq.(\ref{eq:stabilityhom}), the following two conditions are satisfied:
\begin{eqnarray}
    &A&=D_0\partial_v g+D_1\partial_{u} f+(\partial_{\bnd_1v}f-D_{01})(\partial_{\bnd^\top_1u}g-D_{10})>0,\nonumber \\
    &A^2&>4D_0D_1\partial_{u} f \  \partial_v g.
\end{eqnarray}
\section{Derivation of the conditions for the onset of Turing pattern in presence of a cubic Dirac cross-diffusion terms}

\label{appD:conditions_cubic}
\setcounter{equation}{0}
\renewcommand{\theequation}{D\arabic{equation}}
\setcounter{figure}{0}
\renewcommand{\thefigure}{D\arabic{figure}}
In this Appendix we derive the condition for the onset of the Turing instability in presence of a cubic Dirac cross-diffusion term.
The Turing instability is observed when the eigenvalue $\lambda$ satisfying ~\eqref{eq:lambdaXdiff} is positive.
Let us note that the second order equation~\eqref{eq:lambdaXdiff} has both leading coefficients positive. According to Descartes' rule of signs, this equation only admits a positive root if $\Tilde{\Gamma}_2(b_k^2)<0$. Consequently, a positive dispersion on a finite number of modes can be guaranteed by requiring that $\Tilde{\Gamma}_2(b_k^2)<0$ on a finite range of $b_k^2$. \\
First of all, we need to ensure that Eq.~\eqref{eq:lambdaXdiff} admits no positive root in the limit $b_k^2\to \infty$ \textcolor{black}{to avoid long wavelength instability}. This can only be guaranteed if we impose that $D_{01}D_{10}<0$. This ensures that there is a $\Bar{b}_k^2$ such that $\Tilde{\Gamma}_2(b_k^2)>0$ for all $b_k^2>\Bar{b}_k^2$. Note that $D_{01},D_{10}$ are not diffusion coefficients but cross-diffusion coefficients. Indeed they are  coupling each signal with the projection of the signal defined on a different dimension. For this reason we do not need to limit the values of $D_{01}$ and $D_{10}$ to be positive and we can allow negative values while still retaining their physical meaning.\\
Requiring the existence of positive roots in $\lambda$ over a finite range of $b_k^2<\bar{b}_k^2$ can be done again by studying the roots of $\tilde{\Gamma}_2$ using \textit{Descartes' rule of sign}, remembering that the leading and the last coefficients of $\tilde{\Gamma}_2(b_k^2)$ are positive $(+)$. As $\Tilde{\Gamma}_2(b_k^2)>0$ in the limit $b_k^2\to\infty$, we require that $\Tilde{\Gamma}_2(b_k^2)$ admits a positive root in $b_k^2$ to ensure that the system obeys all conditions required for the existence of Turing patterns.\\
A change in sign in the coefficients of $\Tilde{\Gamma}_2(b_k^2)$ is a necessary condition to guarantee this, which leads to patterns of signs  $(+- -+)$, $(+ - ++)$, $(++ - +)$. All these have two sign changes. Hence, applying the rule of sign, $\tilde{\Gamma}_2(b_k^2)$ can admit either 2 or 0 real positive roots. We now need to find a condition to exclude the case of $0$ positive roots.\\
To do so, we start by using the rule of sign for the polynomial of opposite sign, $\tilde{\Gamma}_2(-b_k^2)$, which yields the number of negative rules of $\tilde{\Gamma}_2(b_k^2)$. The possible sign patterns $(+--+)$, $(+-++)$, $(++-+)$ respectively become $(--++)$, $(---+)$, $(-+++)$. These all have a single sign change. Hence, by the rule of sign, $\tilde{\Gamma}_2(-b_k^2)$ has exactly one positive root. Consequently, $\tilde{\Gamma}_2(b_k^2)$ has exactly one negative root, provided the coefficients fall into one of the cases $(+--+)$, $(+-++)$, $(++-+)$.\\
The last condition can be obtained using the cubic discriminant of $\tilde{\Gamma}_2(b_k^2)$. Indeed, in the above three possible cases, we are guaranteed to have a single negative root. There can be $0$ or two positive roots. Guaranteeing two positive roots can be done by imposing that all roots are distinct, \textcolor{black}{and this can be achieved} by setting the discriminant of $\tilde{\Gamma}_2(b^2_k)$ to be positive.\\
Mathematically, this corresponds to first imposing that one of the following two conditions are satisfied
\begin{eqnarray}\label{cond_cross1}
D_0 D_1 +D_{01} \partial_{{\bf B}^\top_1 u} g + D_{10} \partial_{{\bf B_1} v} f < 0,\\
D_1\partial_{u} f+ D_0\partial_v g+\partial_{{\bf B_1}^\top_1 u} g\ \partial_{{\bf B}_1v} f<0, \label{cond_cross1b}
\end{eqnarray}
which constrain the polynomial to one of the above sign patterns, and ensures and the presence of one negative root and either one or two positive roots.
By imposing that the three roots are distinct mathematically through the discriminant of $\tilde{\Gamma}_2(b_k^2)$ leads to the condition
{
\begin{equation}
 2D_0D_1 K_++D_0^2D_1^2 +K_-^2>0,
\label{cond_cross2}   
\end{equation}
where $K_{\pm}$ is given by 
\begin{equation}
K_{\pm}=D_{01}\partial_{{\bf B}_1^\top u}g\pm D_{10}\partial_{{\bf B}_1v}f
\end{equation}
}

\end{document}